\documentclass[prd,eqsecnum, 
nofootinbib,superscriptaddress]{revtex4-1}
\global\arraycolsep=2pt
\usepackage{stmaryrd}
\usepackage{amsmath}
\usepackage{amssymb}
\usepackage{graphicx}
\usepackage{textcomp}
\usepackage{calrsfs}
\usepackage{yfonts}
\usepackage{bm}
\usepackage{xcolor}
\newcommand{\ben}{\begin{equation*}}                                                  
\newcommand{\een}{\end{equation*}}
\newcommand{\bean}{\begin{eqnarray*}}                                                 
\newcommand{\eean}{\end{eqnarray*}}

\newcommand{\nn}{\nonumber}
\newcommand{\be}{\begin{equation}} 
\newcommand{\ee}{\end{equation}}
\newcommand{\bea}{\begin{eqnarray}}
\newcommand{\eea}{\end{eqnarray}}

\DeclareMathOperator{\tr}{tr}

\DeclareMathOperator{\Ai}{Ai}
\DeclareMathOperator{\Bi}{Bi}

\begin{document}
\title{Quantum Electromagnetic Stress Tensor in an Inhomogeneous Medium}
\author{Prachi Parashar}
  \email{prachi.parashar@ntnu.no}
\affiliation{Department of Energy and Process Engineering,
Norwegian University of Science and Technology, 7491 Trondheim, Norway}

\author{Kimball A. Milton}
  \email{kmilton@ou.edu}
  \affiliation{H. L. Dodge Department of Physics and Astronomy,
University of Oklahoma, Norman, OK 73019 USA}

\author{Yang Li}
  \email{liyang@ou.edu}
  \affiliation{H. L. Dodge Department of Physics and Astronomy, University of
Oklahoma, Norman, OK 73019 USA}

\author{Hannah Day}
  \email{Hannah.J.Day-1@ou.edu}
  \affiliation{H. L. Dodge Department of Physics and Astronomy, University of
Oklahoma, Norman, OK 73019 USA}

\author{Xin Guo}
  \email{guoxinmike@ou.edu}
  \affiliation{H. L. Dodge Department of Physics and Astronomy, University of
Oklahoma, Norman, OK 73019 USA}


\author{Stephen A.  Fulling}\email{fulling@math.tamu.edu}
\affiliation{Departments of Mathematics and Physics, Texas A\&M University,
College Station, TX 77843-3368, USA}


\author{In\'es Cavero-Pel\'aez}\email{cavero@unizar.es}
\affiliation{Centro Universitario de la Defensa (CUD), Zaragoza 50090, Spain}

\begin{abstract}
Continuing a program of examining the behavior
of the vacuum expectation value of the stress tensor in
a background which varies only in a single direction, we here study the
electromagnetic stress tensor in a medium with permittivity depending on a
single spatial coordinate, specifically, a planar dielectric
half-space facing a vacuum
region.  There are divergences occurring that are regulated by temporal and
spatial point-splitting, which have a universal character for both 
transverse electric  and transverse magnetic
modes. The nature of the divergences depends on the model of dispersion
adopted.
 And there are singularities occurring at the edge between the dielectric
and vacuum regions, which also have a universal character, depending on the
structure of the discontinuities in the material properties there.  Remarks
are offered concerning renormalization of such models, and the significance of
the stress tensor.  The ambiguity in separating ``bulk'' and ``scattering''
parts of the stress tensor is discussed.
\end{abstract}

\date{\today}
\maketitle

\section{Introduction}
Most studies of the Casimir effect deal with quantum fluctuation forces between
rigid bodies separated by vacuum.  Such forces are finite and can be
calculated exactly, in principle.  (For reviews, see, for example,
\cite{casimirbook,Bordag:2009zzd,Dalvit:lnp}.)  
Casimir's original configuration
was that of perfectly conducting plates in otherwise empty space
\cite{Casimir:1948dh}.  This was generalized by Lifshitz to dielectric slabs,
but again they were separated by vacuum \cite{Lifshitz:1956zz}.
The addition of Dzyaloshinskii and Pitaevskii
was essential to the replacement of the
intervening vacuum by a homogeneous medium \cite{Dzyaloshinskii}.
The resulting theory has been remarkably successful, and was confirmed
by the verification of the attractive force of a helium film by a substrate
\cite{sabisky0,sabisky}, 
well before the modern demonstration of the vacuum Casimir
force \cite{lamoreaux}. The theory has been applied to a wide variety of fields
\cite{Simpson,buhmann,Rodriguez,Mahanty,Parsegian,Krech}.

The local Casimir energy density and other components of the stress tensor 
have also been intensively investigated. These exhibit well-known
behaviors near the surfaces of the bodies.  
(For a review of some of the literature
on this, see Ref.~\cite{Milton:2010qr}.)  This is relevant, not only for
a deeper understanding of the Casimir force, but fundamentally for the
coupling to gravity; in simple contexts, the local Casimir stress tensor
 has been shown to be consistent with the equivalence principle, including
the divergent contributions \cite{Milton:2014psa}.  Consistent results for
finite Casimir stress tensor components
were earlier obtained in Refs.~\cite{Saharian:2003dr,Bimonte:2006dv}.

At least formally, separating rigid  bodies by a uniform dielectric leads
to no difficulties in computing vacuum forces, 
and even dispersion can be incorporated,
although including dissipation may present challenges.  However,
the situation is much less clear when the bodies are immersed in an 
inhomogeneous medium.  
There have been various attempts to describe Casimir forces with nonuniform
dielectrics \cite{Leonhardt:2011zz,Xiong:2013uba,simpson-horsley-leonhardt}.
The most ambitious treatment of the inhomogeneous
electromagnetic Casimir problem seems to be that of Griniasty and Leonhardt
\cite{Griniasty:2017ofc,Griniasty:2017iix}, who examine the local stress tensor
and propose a specific renormalization
scheme to remove the divergences that occur in such circumstances.  
For the case
of a one-dimensional slab with a dielectric response that varies
smoothly except for a discontinuity in the slope as one enters the material, 
they find a universal singularity
behavior in the normal-normal component of the vacuum expectation value of
the stress tensor at the boundary between vacuum and the dielectric.

For some years we have been investigating similar issues, but in the scalar
field context \cite{Bouas:2011ik,Milton:2011iy,Murray:2015tim,Milton:2016sev}.
In particular, using a WKB analysis, we identified the universal Weyl
divergences in the stress tensor components for an arbitrary semi-infinite
 slab described by
a potential $v(z)$, where $z$ is the distance into the slab.  For particular
cases (a linear or a quadratic wall) we also examined how the remainder of the
stress tensor, after the divergent and growing terms are removed, behaves near
the edge.  In this connection the work of  Mazzitelli et al.\ 
should be mentioned \cite{Paz:1988mt,Mazzitelli:2011st}.
(For more references, see the appendix of Ref.~\cite{Murray:2015tim},
and also Ref.~\cite{Bordag1996}, which should have been included there.)
Very recently, we have made further progress in
understanding how the divergences are to be renormalized \cite{fulling2018}.

In the present paper, 
inspired by the remarkable results of Ref.~\cite{Griniasty:2017iix},
we generalize our considerations 
\cite{Bouas:2011ik,Milton:2011iy,Murray:2015tim,Milton:2016sev}
of the local stress tensor in one-dimensional
geometries to the electromagnetic case, in which the role of the potential
is played by the permittivity.  More precisely,
 the deviation of the permittivity from its vacuum value will be referred to
as the potential in this paper.
  In the next section, we review
the difficulty of formulating the stress tensor in inhomogeneous media, and
derive the non-conservation law 
satisfied classically by the spatial stress tensor.
In Sec.~\ref{green} we show how the Green's dyadic for this problem breaks up
into transverse electric (TE) and transverse magnetic (TM) parts.  
We also write
down the construction of the various components of the stress tensor in terms
of the TE and TM Green's functions.  This also includes the correct
dispersive factor for the energy density \cite{Milton:2010yw}.

The generic set-up of the problem is given in Sec.~\ref{sec:gen}, including the
break-up of the Green's functions into ``scattering'' and ``bulk'' parts,
referring to the contributions from the outgoing wave and incoming wave
contributions.  This break-up, of course, is not unique.  An example, the
reflectionless potential considered in Ref.~\cite{Griniasty:2017iix},
is treated somewhat more generally in Sec.~\ref{reflless}. There we show, 
using the
uniform (Debye) asymptotic expansions for the modified Bessel functions,
that there are two types of singularities in the normal-normal component of the
stress tensor  occurring at the edge between the vacuum
and dielectric region: a cubic singularity if there is a discontinuity in the
permittivity, and a quadratic one (coinciding with that found in 
Ref.~\cite{Griniasty:2017iix}) if only the derivative of the permittivity is
discontinuous. We also show that the bulk term 
(the term independent of the reflection coefficient)
contains the expected leading
Weyl divergence, as well as further divergences involving the potential,
 which are regulated by point-splitting.

A second example for which the TE and TM Green's functions may be exactly found
is given in Sec.~\ref{sec:exppot}.  The same edge behavior  is found as in 
Sec.~\ref{reflless} for the continuous case.  This behavior is evidently
universal, as claimed by Ref.~\cite{Griniasty:2017iix}, and we demonstrate
that explicitly in Sec.~\ref{Sec:first-order}, 
using a general perturbative expansion of the
Green's functions.  All of the above neglects dispersion.  
In Sec.~\ref{Sec:disp} we discuss the more realistic
 plasma model, which results 
in the elimination of the edge singularity in the normal-%
normal stress, but yields the divergence structure for the bulk contribution
coinciding with that for the scalar case considered in 
Ref.~\cite{Milton:2016sev}. For the plasma model
of dispersion,
the TE Green's function is identical with  the scalar one. 

Other components of the stress tensor are considered in Sec.~\ref{other}.  
Again, for the plasma model,
the divergences arising from the bulk term in the Green's function coincide
with those found for the scalar situation for both TE and TM modes, and the
edge singularity for the TE mode for the energy density 
coincides with that found for the canonical scalar energy density in 
Ref.~\cite{Milton:2016sev}, while the TM mode has a different numerical
coefficient.

The break-up into bulk and scattering parts is not unique, because
we can always add an arbitrary admixture of the exponentially suppressed
fundamental solution to the exponentially growing one.  We attempt to explore
this further in Sec.~\ref{sec:exlin}, 
for the TE mode, which can be exactly solved for
a potential that depends on the $z$ coordinate linearly.  Numerically, we
show that the scattering part of the energy density and the
normal-normal component of the  stress tensor 
rapidly go to zero as
the dielectric is penetrated, the  former
exhibiting the expected edge singularity.
If an admixture of the first solution is added to the second, the edge
singularities do not change, but the behavior inside the dielectric is 
altered, but still
tending to zero as one goes deeply within the material.  Only if the
scattering part of the Green's function is completely suppressed (a set of 
measure zero in parameter space) does the qualitative (and quantitative, 
for the divergences and edge singularities) behavior change.

We finally consider a situation with mirror symmetry in Sec.~\ref{sec:mirror}.
Here we consider two reflected 
potentials meeting at $z=0$ so there is no vacuum
region.  In this case, not surprisingly, the edge singularity is doubled.
Concluding remarks are offered in Sec.~\ref{sec:concl}.  In Appendix 
\ref{appa} we explain the point-split regulation we use in this paper, while
in Appendix \ref{appb} we develop the perturbation theory for a potential
which is both continuous and has a continuous first derivative, but where
the second derivative is discontinuous. 

In this paper we use Heaviside-Lorentz electromagnetic units, and $\hbar=c=1$.

\section{Force on dielectric}
\label{sec:force}
From the Maxwell-Heaviside 
equations we can derive the statement of electromagnetic
momentum conservation.  
We follow Sec.~7.1 of Ref.~\cite{ce}.  Equation (7.10) there says that
\be
\mathbf{f}+\frac\partial{\partial t}\mathbf G=
-D_i\bm{\nabla}E_i+\bm{\nabla}\cdot(\mathbf{DE})-B_i\bm{\nabla}H_i+\bm{\nabla}
\cdot(\mathbf{BH}),
\label{pcons}
\ee
where
\be \mathbf{f}=\rho\mathbf{E}+\mathbf{j\times B} \ee
is the force density on the charged particles, and the field momentum is
\be\mathbf{G}=\mathbf{D\times B}.\ee
Here, a summation convention is used for repeated indices, and
 $\rho$ and $\mathbf{j}$ are the free charge and current densities.
To what extent is the right side of Eq.~(\ref{pcons}) the negative of a total 
divergence,  $-\bm{\nabla}\cdot\mathbf{T}$, which would imply a local
conservation law of momentum?
As usual it is convenient to do a Fourier (frequency) transform of the
fields (we will here suppress the spatial coordinates), assuming a
linear medium.  For the electric fields
\be
\mathbf{E}(t)=\int\frac{d\omega}{2\pi} e^{-i\omega t}\mathbf{E}(\omega),\quad 
\mathbf{D}(t)=\int\frac{d\omega}{2\pi} e^{-i\omega t}\bm{\varepsilon}(\omega)
\cdot\mathbf{E}(\omega),
\ee
where we have introduced a frequency-dependent permittivity tensor, 
$\bm{\varepsilon}(\omega)$,
which we allow to be spatially varying.  Similarly for the magnetic
fields,
\be
\mathbf{H}(t)=\int\frac{d\omega}{2\pi} e^{-i\omega t}\mathbf{H}(\omega),\quad 
\mathbf{B}(t)=\int\frac{d\omega}{2\pi} e^{-i\omega t}\bm{\mu}(\omega)
\cdot\mathbf{H}(\omega).
\ee
We now take the average over a time $T$ large compared to atomic time scales
but short compared to macroscopic times, so the dyadic product can be
written, for example, as
\be
\overline{\mathbf{D}(t)\mathbf{E}(t)}=\frac1T \int\frac{d\omega}{2\pi}[
\bm{\varepsilon}(\omega)\cdot\mathbf{E}(\omega)]\mathbf{E}(\omega)^*.
\ee
Then, in the absence of dissipation,  we use the Hermiticity property
arising from the reality of the constitutive relations in spacetime, 
$\varepsilon_{ij}(\omega)=\varepsilon_{ji}(-\omega)=
\varepsilon_{ji}(\omega)^*$.\footnote{That is, $\bm{\varepsilon}^\dagger
=\bm{\varepsilon}$.  This cannot be true if dissipation is present.  In
that case, if we suppose $\bm{\varepsilon}$ is symmetric, 
$\Re \bm{\varepsilon}$ and $\Im \bm{\varepsilon}$ are then
both diagonalizable, but in different bases.}
If the permittivity and permeability were independent of position, there would 
be an averaged macroscopic stress tensor,
\be
\overline{\mathbf{T}}=\frac1T\int\frac{d\omega}{2\pi}\left(
\frac{\bm{1}}2[\mathbf{D(\omega)^*\cdot E(\omega)+B(\omega)^*\cdot H(\omega)}]
-\mathbf{D(\omega)^*E(\omega)-B(\omega)^*H(\omega)}\right).
\ee
However,  if the electrical 
properties depend on position, this is not the case, but,
rather,  the right side of Eq.~(\ref{pcons}) would be
\be
-\overline{\bm{\nabla}\cdot \mathbf{T}}+\frac1{2T} 
\int\frac{d\omega}{2\pi}\left[
E_i(\omega)^* (\bm{\nabla}\varepsilon_{ij}(\omega))E_j(\omega)
+H_i(\omega)^* (\bm{\nabla}\mu_{ij}(\omega))H_j(\omega)\right].\label{divtn0}
\ee
For a recent review concerning electromagnetic
stress tensors see Ref.~\cite{brevik17}.

 For example,
 consider a dielectric body ($\bm{\mu}=\bm{1}$) immersed in a static
 classically imposed electric field.
Because there is no time dependence and no free charge,  we have
\be
\bm{\nabla}\cdot \mathbf{T}=\frac1{2}
\tr\mathbf{ E E}(\bm{\nabla}) 
\bm{\varepsilon},
\ee
where the trace is over the tensor indices, and the 
notation $(\bm{\nabla})$ is a reminder that the free vector index is on the
gradient operator.  Suppose the body, which need not be homogeneous, is 
immersed
in a homogeneous medium of permittivity $\bm{\epsilon}$.  The force on the body
is the momentum flux into the body,
\be
\mathbf{F}=-\oint_{S} d\mathbf{S}\cdot \mathbf {T},
\ee
since the local momentum conservation law holds there, where $S$ is a surface 
that entirely surrounds the body.  By the divergence theorem
\be
\mathbf{F}=-\int_V (d\mathbf{r}) \bm{\nabla}\cdot\mathbf{T}=
-\frac1{2}\int_V(d\mathbf{r}) \tr
\mathbf{E E} (\bm{\nabla})\bm{\varepsilon},
\ee
where the spatial integral is over the interior 
of the body (because the permittivity is constant outside the body).
This is a generalization of the familiar formula for the force on a
dielectric, Eq.~(11.44) of Ref.~\cite{ce}, to which it reduces for the
isotropic case.

We can immediately generalize this to the Casimir force by replacing in 
Eq.~(\ref{divtn0})
\be
\langle\mathbf{ E(\omega) E(\omega')^*}\rangle=2\pi\delta(\omega-\omega')
\frac1i\bm{\Gamma}(\omega), 
\ee
in terms of the Green's dyadic $\bm{\Gamma}$,
so that the dispersion force on the dielectric body is
\be
\mathbf{F}_{\rm Cas}= -\frac1{2i} \int(d\mathbf{r})\int\frac{d\omega}{2\pi}
\tr \bm{\Gamma}(\mathbf{r,r;\omega})(\bm{\nabla})\bm{\varepsilon}(\mathbf{r},
\omega).
\ee  Here we have identified $2\pi\delta(0)$ with the averaging time $T$.
In particular, if the body has a homogeneous dielectric constant 
$\bm{\varepsilon}\ne\bm{\epsilon}$, then
\be
\bm{\nabla}\bm{\varepsilon}=-\mathbf{\hat s}(\bm{\varepsilon-\epsilon})
\delta(s-s_0(\mathbf{r_\perp})),
\ee
where the surface of the body is given by $s=s_0(\mathbf{r_\perp})$, 
in terms of
a coordinate $s$ (outwardly) normal to the surface. The other coordinates
are denoted by $\mathbf{r_\perp}$.  
(For the case of a planar body in the $x$-$y$ plane, $s=z$.)
Thus the Casimir force on the body is given by an integral over the
surface of the body,
\be
\mathbf{F}_{\rm Cas}=\frac1{2i}\oint_S d\mathbf{S} \int\frac{d\omega}{2\pi} 
\tr\bm{(\varepsilon-\epsilon)}(\mathbf{r},\omega)
\bm{\Gamma}(\mathbf{r,r};\omega).
\ee
Again, this is an obvious generalization of known formulas.\footnote{For 
example, 
for the case
of a dielectric ball, this formula leads immediately, upon use of the 
orthogonality relations for the vector spherical harmonics given in 
Ref.~\cite{ce},  p.~534, to the expression (5.19)
for the total outward stress given  in Ref.~\cite{casimirbook}.}
The general form for the nonconservation of the vacuum expectation
value of the electromagnetic stress tensor in a medium is
\be
\overline{\bm{\nabla}\cdot\langle\mathbf{T}(\mathbf{r})\rangle}=
\frac1{2i}\int\frac{d\omega}{2\pi} \tr \bm{\Gamma}(\mathbf{r,r};\omega)
(\bm{\nabla})\bm{\varepsilon}(\mathbf{r},\omega),\quad 
\mbox{or}\quad \overline{\partial_j \langle T_{ji}\rangle
(\mathbf{r})}
=\frac1{2i}\int\frac{d\omega}{2\pi}\Gamma_{jk}(\mathbf{r,r};\omega)
\partial_i\varepsilon_{kj}(\mathbf{r},\omega).
\ee
This is, of course, quite analogous to the nonconservation equation 
satisfied by the stress tensor for
 a scalar field in a background potential \cite{Milton:2016sev}.

\section{Green's Functions}
\label{green}
In this paper we will consider planar situations in which the permittivity
$\varepsilon(z)$ and the permeability $\mu(z)$
depend only on a single coordinate $z$.  
We will also allow $\varepsilon$ and $\mu$ to depend on 
frequency. For simplicity, we will henceforth assume that $\varepsilon$
and $\mu$ are isotropic. It is also convenient to make a Euclidean
transformation $\omega\to i\zeta$.
 The general Green's dyadic obeys an equation which follows from the 
Maxwell-Heaviside equations, 
\be
\left(-\frac1{\zeta^2}\bm{\nabla}\times\frac1{\mu}\bm{\nabla}\times-\varepsilon
\bm{1}\right)\cdot\bm{\Gamma}=\bm{1},
\ee which breaks into two modes, TE and TM modes, denoted by two scalar Green's
functions labelled by E and H, respectively.  These satisfy
the differential equations
\begin{subequations}
\label{ehgfeqn}
\bea
\left(-\frac\partial{\partial z}\frac1\mu\frac\partial{\partial z}+\frac{k^2}
\mu+\zeta^2\varepsilon\right)g^E(z,z')&=&\delta(z-z'),\\
\left(-\frac\partial{\partial z}\frac1\varepsilon\frac\partial{\partial z}
+\frac{k^2}\varepsilon
+\zeta^2\mu\right)g^H(z,z')&=&\delta(z-z').
\eea
\end{subequations}
The spatial Fourier components of $\bm{\Gamma}$, defined by
\be
\bm{\Gamma}(\mathbf{r,r'})
=\int\frac{(d\mathbf{k}_\perp)}{(2\pi)^2}e^{i\mathbf{
k_\perp\cdot (r-r')_\perp}} \bm{\gamma}(z,z'),
\ee
are given in terms of these two scalar Green's functions, in the coordinate
system where $\mathbf{k}_\perp$ has only a component in the $x$ direction
(we drop the $z$, $z'$ dependence of $g^E$ and $g^H$):
\be
\bm{\gamma}(z,z')=\left(\begin{array}{ccc}
\frac1{\varepsilon}\partial_z\frac1{\varepsilon'}
\partial_{z'}g^H-\frac1{\varepsilon}\delta(z-z')
&0&\frac{ik}{\varepsilon\varepsilon'}\partial_z g^H\\
0&-\zeta^2g^E&0\\
-\frac{ik}{\varepsilon\varepsilon'}\partial_{z'}g^H&0&\frac{k^2}{\varepsilon
\varepsilon'}g^H-\frac1{\varepsilon}\delta(z-z')
\end{array}\right).
\ee
Here $\varepsilon=\varepsilon(z)$, $\varepsilon'=\varepsilon(z')$.
These are just as given in Refs.~\cite{casimirbook,Schwinger:1977pa}.

The Fourier-transformed electromagnetic stress tensor may also be given
in simple form in terms of these two scalar Green's functions.  For example,
the $zz$ component of the reduced stress tensor is simply
\be
t_{zz}(z)=\frac1{2\varepsilon'}
\left[\partial_z\partial_{z'}-(k^2+\zeta^2\varepsilon'\mu)
\right]g^H+
\frac1{2\mu'}\left[\partial_z\partial_{z'}-(k^2+\zeta^2\varepsilon\mu')
\right]g^E,\label{stform}
\ee
where after differentiation, the limit $z\to z'$ is understood.

Let us also record the other diagonal components of the reduced stress tensor.
First,  the energy density, which must include the dispersive factors:
\be
t_{00}(z)=\frac1{2}\frac{d(\omega\varepsilon)}{d\omega}\left(
\frac1\varepsilon\partial_z\frac1{\varepsilon'}\partial_{z'}g^H-\zeta^2g^E
+\frac{k^2}{\varepsilon\varepsilon'}g^H\right)+
\frac1{2}\frac{d(\omega\mu)}{d\omega}
\left(\frac1\mu\partial_z\frac1{\mu'}\partial_{z'}g^E-\zeta^2g^H+\frac{k^2}{\mu
\mu'}g^E\right).\label{endenform}
\ee
To preserve the symmetry between the transverse components 
of the reduced stress tensor,
 we rotate $\bm{\gamma}$ to a general
coordinate system.  Doing so does not affect $t_{00}$ and $t_{zz}$, but yields
after using the equations of motion (\ref{ehgfeqn})
\begin{subequations}
\bea
t_{xx}(z)&=&\frac1{2\varepsilon'}
\left[-\frac{k_x^2-k_y^2}{k^2}\left(\partial_z\partial_{z'}+\zeta^2\varepsilon'
\mu\right)+k^2\right]g^H+
\frac1{2\mu'}\left[-\frac{k_x^2-k_y^2}{k^2}\left(\partial_z\partial_{z'}
+\zeta^2
\varepsilon\mu'\right)+k^2\right]g^E,\label{stxxform}\\
t_{yy}(z)&=&\frac1{2\varepsilon'}
\left[-\frac{k_y^2-k_x^2}{k^2}\left(\partial_z\partial_{z}'+\zeta^2
\varepsilon'\mu\right)+k^2\right]g^H+
\frac1{2\mu'}\left[-\frac{k_y^2-k_x^2}{k^2}\left(\partial_z\partial_{z'}
+\zeta^2\varepsilon\mu'\right)+k^2\right]g^E.\label{styyform}
\eea
\end{subequations}
There are also off-diagonal terms, linear in $k_x$ or $k_y$, 
which would vanish upon regulated integration,
if that regulation respects the two-dimensional rotational symmetry of the
problem. Such a regulator reduces $t_{xx}$ and $t_{yy}$ to
\be
t_{xx}=t_{yy}=\frac{k^2}2\left(\frac1\varepsilon g^H+\frac1\mu g^E\right).
\label{txxtyy}
\ee
The four-dimensional trace
\be
t^\mu_\mu=t_{zz}+t_{xx}+t_{yy}-t_{00}=-\frac12\frac\omega\varepsilon
\frac{d\varepsilon}{d\omega}\left[\frac1{\varepsilon'}(\partial_z\partial_{z'}
+k^2)g^H-\zeta^2\varepsilon g^E\right]
-\frac12\frac\omega\mu
\frac{d\mu}{d\omega}\left[\frac1{\mu'}(\partial_z\partial_{z'}
+k^2)g^E-\zeta^2\mu g^H\right]\label{traceid}
\ee
is zero if there is no dispersion.

\section{Generic planar problem}
\label{sec:gen}
To save typographical space, we use comma-separated notation, $(\mu,
\varepsilon)$ and $(E,H)$ to write the TE and TM mode expressions in the 
following.
We can construct the Green's functions from the solutions of the homogeneous
equations
\be
\left[-\partial_z\frac1{\mu,\varepsilon}\partial_z+\frac{k^2}
{\mu,\varepsilon}+\zeta^2(\varepsilon,\mu)\right]\left\{\begin{array}c
F^{E,H}\\G^{E,H}\end{array}\right.=0.\label{fgdeq}
\ee
Here we take $F$ to denote a solution that does not diverge for $z\to\infty$
(typically goes to zero),
while $G$ is an arbitrary independent solution.  The Wronskian of these
two solutions is
\be
w(z)=F(z)G'(z)-G(z)F'(z).
\ee
We want to solve the Green's function equations (\ref{ehgfeqn}) in terms
of these solutions, for the situation of a ``soft wall'', where
\be
\mu(z), \varepsilon(z)=\left\{\begin{array}{cc}
1,&z<0,\\
\tilde{\mu}(z), \tilde\varepsilon(z),&z>0.\end{array}\right.
\ee
The  solutions are ($\kappa=\sqrt{k^2+\zeta^2}$)
\be
g^{E,H}(z,z')=\left\{\begin{array}{cc}
\frac1{2\kappa}\left[e^{-\kappa|z-z'|}+R^{E,H}e^{\kappa(z+z')}\right],
&z,z'<0,\\
\frac1{\alpha^{E,H}}\left[F^{E,H}(z_>)G^{E,H}(z_<)+\tilde R^{E,H}F^{E,H}(z)
F^{E,H}(z')\right],&z,z'>0.\end{array}\right.\label{hegf}
\ee
Here, the constant $\alpha$ is related to the Wronskian by
\be
\alpha^{E,H}=\frac{w^{E,H}(z)}{\tilde{\mu}(z),\tilde{\varepsilon}(z)}.
\label{effw}
\ee
The reflection coefficients are determined by requiring that $g^{E.H}$ be
continuous at $z=0$, and that $\frac1{\mu,\varepsilon}\partial_z g^{E,H}$ 
also be
continuous there.  This corresponds to the continuity of 
$\hat{\mathbf z}\times {\mathbf E}$ and $\hat{\mathbf z}\cdot \mathbf{B}$,
and of
$\hat{\mathbf z}\times {\mathbf H}$ and $\hat{\mathbf z}\cdot \mathbf{D}$.
(Imposing these matching conditions 
 requires the form of the Green's function for
$z_> >0>z_<$, not displayed here.)  The consequence is
\be
R^{E,H}=\frac{\kappa F^{E,H}(0)+\frac1{\mu,\epsilon}F^{E,H\,\prime}(0)}
{\kappa F^{E,H}(0)-\frac1{\mu,\epsilon}F^{E,H\,\prime}(0)}
\ee
and
\be
\tilde R^{E,H}=-\frac{\kappa G^{E,H}(0)-\frac1{\mu,\epsilon}
G^{E,H\,\prime}(0)}
{\kappa F^{E,H}(0)-\frac1{\mu,\epsilon}F^{E,H\,\prime}(0)}.\label{intrc}
\ee
Here $\mu=\tilde\mu(0)$, $\epsilon=\tilde\varepsilon(0)$.

In the above construction, $G$ is completely arbitrary, save that it
be a solution, independent of $F$, to the differential equation (\ref{fgdeq}).
Therefore, the reflection coefficient $\tilde R$ is not unique, and indeed
can be made equal to zero by the replacement $G\to G-\tilde R F$.  To have
a unique reflection coefficient, we need a condition to determine the form
of $G$.  Such is supplied by imposing a boundary condition at $z\to-\infty$,
even though this is outside the region $z,z'>0$ where the construction
(\ref{hegf}) holds.  That is, assuming the continuous functions $\tilde
\varepsilon(z)$, $\tilde\mu(z)$ hold in all space,  so there is
no discontinuity,  we will
henceforth choose $G$ subject to the boundary condition
\be
z\to-\infty:\quad G^{E,H}\to 0.
\ee
Then the reflection coefficient is uniquely defined.
(These boundary conditions as stated here are somewhat schematic;
the specific conditions at $\pm\infty$ depend on the structure of 
$\varepsilon(z)$.)

The stress in the vacuum region, to the left of the wall ($z<0$), is
immediately calculated from Eq.~(\ref{stform}):
\be
z<0:\quad t_{zz}^{E,H}=-\frac\kappa2,\label{outstress}
\ee
which is independent of $z$, the term involving the reflection coefficient
having cancelled out.  This is universally recognized as an irrelevant
bulk term, since it has no contribution from the wall, and would be present
if vacuum filled all space, so is to be omitted.

It is the assertion of Ref.~\cite{Griniasty:2017iix} that the same
omission is to be done for the contribution to the stress tensor coming
from the part of the Green's function in the $z>0$ region
 that is not proportional
to the reflection coefficient: In particular, they advocate omitting the stress
tensor contribution arising from the 
term in the Green's function (\ref{hegf}) $\frac1\alpha F(z_>)G(z_<)$,
even though it is spatially varying, because this term would be there
in the absence of the edge at $z=0$.  
This hypothesis may be suspect, but
we will follow it for the moment. 

\section{Exactly solvable examples}
\label{sec:exactex}
Now we examine two cases where both the TE and TM modes may be explicitly
given. In the first example, the permittivity
has a singularity at a finite value of $z$,  which is the
natural boundary of the problem,
and in the second the permittivity has an exponential behavior.
\subsection{A first example}
\label{reflless}
Let us consider a planar medium described by
\be
a>z>0:\quad \tilde\mu=1,\quad \tilde\varepsilon(z)=\frac\lambda{(a-z)^2},
\ee
which has a singularity at $z=a$.  (This is a slightly generalized
version of the medium considered in Ref.~\cite{Griniasty:2017iix}, where
the potential was continuous, so
$\epsilon\equiv\tilde\varepsilon(0)=\lambda/a^2=1$.)  
Because of that singularity,
the right side of the wall has a finite depth, $0<z<a$;  the region $z>a$ 
is completely disconnected from the region containing the wall.  This potential
has the virtue of allowing explicit solutions:
\be
F^{E,H}=(a-z)^{\pm1/2}I_\nu(k(a-z)),\quad G^{E,H}=(a-z)^{\pm1/2}K_\nu(k(a-z)),
\ee
where
\be
\nu=\sqrt{\lambda\zeta^2+\frac14},\quad \frac1{\alpha^{E,H}}=1,\lambda.
\ee
Here $F$ is chosen to be finite as $z\to a$. 
Indeed, $I_\nu(0)=0$, $K_\nu(+\infty)=0$, consistent with the criteria
stated in the previous section.
Then the reflection coefficients in the medium are
\be
\tilde R^{E,H}=-\frac{\frac{ka}{1,\epsilon}K_\nu'(ka)+\left(\kappa a
\pm\frac1{2(1,\epsilon)}\right)K_\nu(ka)}
{\frac{ka}{1,\epsilon}I_\nu'(ka)+\left(\kappa a
\pm\frac1{2(1,\epsilon)}\right)I_\nu(ka)}.
\ee
The scattering  part of the $zz$ component of the reduced stress tensor 
(the part proportional to the reflection coefficients) is
\be
 t_{zz}^{s,E,H}(z)=\frac12\tilde R^{E,H}\left\{-\left[k^2(a-z)+\frac{\lambda 
\zeta^2-1/4}{a-z}\right]I_\nu^2(k(a-z))+k^2(a-z)I_\nu^{\prime2}(k(a-z))
\pm k I_\nu(k(a-z))I_\nu'(k(a-z))\right\}.\label{tzzrefl}
\ee

\begin{figure}
\includegraphics{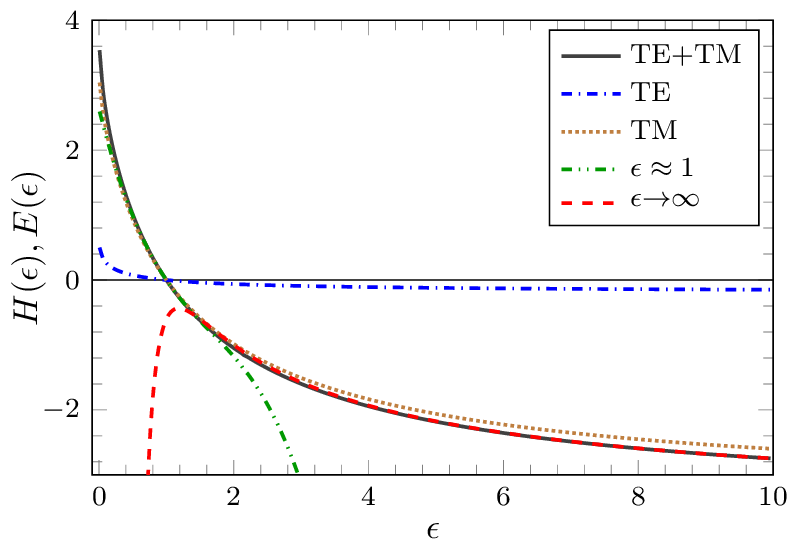}
\caption{\label{fig1} The $\epsilon$-dependent factors in the $zz$ components
of the stress tensor in Eq.~(\ref{eandh}). 
The small and large $\epsilon-1$ limits
go out to third order and $-7/2$ order, respectively. Clearly, the TE 
contribution is almost insignificant, and the two asymptotic limits
accurately cover the full range of $\epsilon$. }
\end{figure}

As we wish to examine the stress just inside the wall,
we can use the uniform asymptotic expansion (UAE) for the Bessel functions,
because it captures the short-distance behavior \cite{nist}. That expansion
is, as $\nu\to\infty$:
\begin{subequations}
\bea
I_\nu(\nu Z)&\sim&\frac1{\sqrt{2\pi\nu}}\frac{e^{\nu\eta(Z)}}{(1+Z^2)^{1/4}}
\left(1+\sum_{k=1}^\infty \frac{u_k(t)}{\nu^k}\right),\quad
K_\nu(\nu Z)\sim\sqrt{\frac\pi{2\nu}}\frac{e^{-\nu\eta(Z)}}{(1+Z^2)^{1/4}}
\left(1+\sum_{k=1}^\infty (-1)^k\frac{u_k(t)}{\nu^k}\right),\\
I'_\nu(\nu Z)&\sim&\frac1{\sqrt{2\pi\nu}}e^{\nu\eta(Z)}\frac{(1+Z^2)^{1/4}}Z
\left(1+\sum_{k=1}^\infty \frac{v_k(t)}{\nu^k}\right),\quad
K'_\nu(\nu Z)\sim-\sqrt{\frac\pi{2\nu}}e^{-\nu\eta(Z)}\frac{(1+Z^2)^{1/4}}Z
\left(1+\sum_{k=1}^\infty (-1)^k\frac{v_k(t)}{\nu^k}\right),\nn\\
\eea
\end{subequations}
where $u_k$ and $v_k$ are polynomials in $t=(1+Z^2)^{-1/2}$.  The first
of these are
\be
u_1(t)=\frac1{24}(3t-5t^3),\quad v_1(t)=\frac1{24}(-9t+7t^3).
\ee
All we need to know about the functions in the exponents is the derivative:
\be
\eta'(Z)=\frac1{Zt}.\label{dereta}
\ee
If we retain only the leading factor in the UAE the reflection coefficients
are approximately
\be
\tilde R^{E,H}\sim -\pi e^{-2\nu\eta(ka/\nu)}\frac{\kappa a\pm
\frac1{2(1,\epsilon)}-\frac1{1,\epsilon}\sqrt{\tilde\kappa^2a^2+1/4}}
{\kappa a\pm
\frac1{2(1,\epsilon)}+\frac1{1,\epsilon}\sqrt{\tilde\kappa^2a^2+1/4}}.
\label{rhel}
\ee
Here $\tilde\kappa^2=\epsilon\zeta^2+k^2$.  In the remaining factor
of Eq.~(\ref{tzzrefl}) we must keep the $O(1/\nu)$ corrections because
they are of the same order in $\kappa a$ as the leading term 
in the stress tensor construction,  leaving for
the rest of the $zz$ component of the reduced stress tensor
\be
 t_{zz}^{s,E,H}\sim \frac{\tilde R^{E,H}}{4\pi (a-z)}e^{2\nu\eta(k(a-z)/\nu)}
\left(\frac{1+4(\tilde\kappa a)^2[v_1(t)-u_1(t)]/\nu}
{2\sqrt{\lambda\zeta^2+k^2(a-z)^2+1/4}}\pm1\right),
\quad t=\left[1+\left(\frac{k(a-z)}{\nu}\right)^2\right]^{-1/2},\label{tzz1uae}
\ee
where $u_1(t)-v_1(t)=\frac{t}2(1-t^2)$. 
The UAE presumes that the significant values of $\tilde\kappa$ are large.
If $\epsilon=\varepsilon(0)\ne1$, in the first approximation we may neglect
terms of order $1/(\tilde\kappa a)$ and smaller, so the reflection coefficients
reduce to
\be
\tilde R^{E,H}\approx -\pi e^{-2\nu\eta(ka/\nu)}\frac{\kappa-\frac1{1,\epsilon}
\tilde\kappa}{\kappa+\frac1{1,\epsilon}
\tilde\kappa},\label{rcdiscepsilon}
\ee
which has the form familiar from a step discontinuity in the dielectric
constant.  Further, near the boundary, the exponents combine:
\be
2\nu\eta(k(a-z)/\nu)-2\nu \eta(ka/\nu)\approx -2\tilde \kappa z,
\ee
which makes use of Eq.~(\ref{dereta}).  Finally, to carry out the integrals 
over frequency and transverse wavevectors we adopt polar coordinates, so that
\be
\int d\zeta\int (d\mathbf{k}_\perp)=
\frac1{\sqrt{\epsilon}}\int_0^\infty d\tilde\kappa\,\tilde\kappa^2
\int_{-1}^1 d\cos\theta\int_0^{2\pi}d\phi,\label{polarcoords}
\ee
with $\sqrt{\epsilon}\zeta=\tilde\kappa\cos\theta$, $k=\tilde\kappa\sin\theta$.
The angle $\theta$  occurs inside the two reflection coefficients,
as well as inside the formula for $t^s_{zz}$, Eq.~(\ref{tzz1uae}),
since near the wall $[u_1(ka/\nu)-v_1(ka/\nu)]/\nu=(1-\cos^2\theta)/
(2\tilde\kappa a)$,
and the integrals of these dependencies for the TE and TM modes give
\begin{subequations}
\label{eandh}
\bea
E(\epsilon)&=&\int_{-1}^1 d\cos\theta\cos^2\theta
\frac{\sqrt{(1/\epsilon-1)\cos^2\theta+1}-1}
{\sqrt{(1/\epsilon-1)\cos^2\theta+1}+1},\\
H(\epsilon)&=&\int_{-1}^1 d\cos\theta\,(\cos^2\theta-2)
\frac{\sqrt{(1/\epsilon-1)\cos^2\theta+1}-1/\epsilon}
{\sqrt{(1/\epsilon-1)\cos^2\theta+1}+1/\epsilon}.
\eea
\end{subequations}
The functions $E(\epsilon)$ and $H(\epsilon)$ are elementary, given in
terms of logarithms, but are not very illuminating to display.
Instead we show the plot of them in Fig.~\ref{fig1}, and give the limits
for small and large values of $\epsilon-1$:
\begin{subequations}
\bea
\epsilon-1\ll1:&&\quad E(\epsilon)\sim -\frac1{10}(\epsilon-1)
+\frac9{140}(\epsilon-1)^2+\dots,\quad
H(\epsilon)\sim-\frac{43}{30}(\epsilon-1)+\frac{93}{140}
(\epsilon-1)^2+\dots,\\
\epsilon\gg1:&&\quad 
E(\epsilon)\sim \pi-\frac{10}3+\frac{3\pi/2-4}{\epsilon}
-\frac43\frac1{\epsilon^{3/2}}+\dots,\quad
H(\epsilon)\sim-\frac{10}3+\frac{3\pi}\epsilon-\frac4{\epsilon^{3/2}}+\dots.
\eea
\end{subequations}
The remaining integral on $\tilde\kappa$ is simple,  so 
after integrating over $\mathbf{k}$ and $\zeta$, we are left
with the ``scattering part'' of the $zz$ component of the
stress tensor  near the wall ($z\to0+$):
\be
T_{zz}^{s,E,H}\sim -\frac1{64\pi^2\sqrt\epsilon}\frac1{az^3}
\left\{\begin{array}{c}
E(\epsilon)\\H(\epsilon)
\end{array}\right..\label{theepsdisc}
\ee
And the total $zz$ component of the stress is the sum of these two components,
which for the case of a small discontinuity reduces to 
\be
 T_{zz}^{s,E+H}=\frac{23}{960\pi^2}
(\epsilon-1)\frac1{az^3}, \quad \frac{E}H=\frac3{43},\quad \epsilon-1\ll1.\label{tzzweak}
\ee

This cubic singularity disappears if there is no discontinuity, that is,
 $\epsilon=1$, where $\tilde\kappa=\kappa$.
  Then we need to keep the order $1/\nu$ correction in the reflection
coefficients as well, so Eq.~(\ref{rhel}) gets modified to
\be
\tilde R^{E,H}\sim\mp\frac{\pi}{4\kappa a}\left[1\mp(1-\cos^2\theta)\right]
e^{-2\nu\eta(ka/\nu)},\label{r0uae}
\ee
which when inserted into Eq.~(\ref{tzz1uae}) yields immediately
\be
t_{zz}^{s, E,H}=-\frac1{16\kappa a^2}[1\mp(1-\cos^2\theta)]^2 e^{-2\kappa z}.
\label{tsehmod1}
\ee
When the integrals over $\kappa$ and $\theta$ are carried out, we obtain
\be
T_{zz}^{s,E,H}\sim -\frac1{1920\pi^2 a^2z^2}\left\{\begin{array}{c}
3\\43\end{array}\right.,\label{edgeex}
\ee
the sum of the two contributions being
\be
T_{zz}^{s,E+H}=-\frac{23}{960}\frac{1}{\pi^2a^2z^2},\label{tzzgl}
\ee
which is exactly the result found in 
Ref.~\cite{Griniasty:2017iix}.  The similarity of the coefficients in
Eqs.~(\ref{tzzweak}) and (\ref{edgeex}), (\ref{tzzgl}) is striking.

We close this subsection by examining the omitted contribution from the 
``bulk'' term in the interior,
\be
 g^{b,E,H}(z,z')=\frac1{\alpha^{E,H}}F^{E,H}(z_>)G^{E,H}(z_<)=(1,\lambda)
(a-z)^{\pm1/2}(a-z')^{\pm1/2}I_\nu(k(a-z_>))K_\nu(k(a-z_<)).
\ee
It is quite obvious that this does not give singular behavior near the
discontinuity in {\color{red}$\varepsilon(z)$ at $z=0$}, 
but it does yield divergent contributions.
The corresponding reduced stress tensor has a form similar to that given in 
Eq.~(\ref{tzzrefl}):
\bea
 t_{zz}^{b,E,H}(z)&=&\frac12\bigg\{-\left[k^2(a-z)+\frac{\lambda 
\zeta^2-1/4}{a-z}\right]I_\nu(k(a-z))K_\nu(k(a-z))
+k^2(a-z)I_\nu^{\prime}(k(a-z))K'_\nu(k(a-z))\nn\\
&&\quad
\mbox{}\pm k \left[I_\nu(k(a-z))K_\nu'(k(a-z))+I_\nu^{\prime}(k(a-z))
K_\nu(k(a-z))\right]\bigg\}.\label{tzzbulk}
\eea
Now, when the UAE is inserted, the cancellation observed in the reflection-%
dependent part does not occur, so the leading term is
\be
t^{b,E,H}_{zz}\sim 
\frac1{4(a-z)}\left(-2\sqrt{\lambda\zeta^2+k^2(a-z)^2}\pm1\right).
\label{tbHEzz}
\ee
For the moment we examine {\color{red} only} the leading term in the limit of
this expression as $z\to0$, which is
\be t^{b,E,H}_{zz}\to-\frac{\tilde\kappa}2,\label{redtsbulk}
\ee
 the obvious generalization of
Eq.~(\ref{outstress}). When this is integrated over all frequencies and
wavenumbers, and regulated by point-splitting as in Ref.~\cite{Milton:2016sev},
we obtain (see Appendix \ref{appa})
\be
T_{zz}^{b,E,H}\sim-\frac1{4\pi^2\sqrt{\epsilon}}
\int_0^\infty d\tilde\kappa\,\tilde\kappa^3
\frac{\sin\tilde\kappa\delta}{\tilde\kappa\delta}
=\frac1{2\pi^2\sqrt{\epsilon}\delta^4},\quad \delta\to0,
\label{tsbulk}
\ee
exactly the leading bulk divergence seen for each scalar mode in 
Ref.~\cite{Milton:2016sev}, apart from the expected index of refraction
factor. Later we shall encounter the subleading divergences dependent on
the potential; beyond them, in the exact $t^{b,E,H}$ there are finite
terms that presumably have physical significance.  

\subsection{Exponential permittivity}
\label{sec:exppot}
Let us give another exactly solvable model.  Consider the permittivity
function
\be
\varepsilon(z)=\left\{\begin{array}{cc}
1,&z<0,\\
e^{\alpha z},&z>0.
\end{array}\right.
\ee
For the two modes, the two fundamental solutions to Eq.~(\ref{fgdeq}) are for
$z>0$ \cite{GR}
\begin{subequations}
\bea
\left.\begin{array}{c}
F^E(z)\\G^E(z)\end{array}\right\}&=&\left\{
\begin{array}{c}K_\nu(2\zeta e^{\alpha z/2}/\alpha),\\
I_\nu(2\zeta e^{\alpha z/2}/\alpha),\end{array}\right.\\
\left.\begin{array}{c}
F^H(z)\\G^H(z)\end{array}\right\}&=&e^{\alpha z/2}\left\{
\begin{array}{c}K_{\tilde\nu}(2\zeta e^{\alpha z/2}/\alpha),\\
I_{\tilde\nu}(2\zeta e^{\alpha z/2}/\alpha),\end{array}\right.
\eea
\end{subequations}
where
\be
\nu=\frac{2k}\alpha,\quad \tilde\nu=\sqrt{1+\frac{4k^2}{\alpha^2}}.
\ee
 Again, the second solution is unique, according to the
criteria enunciated in Sec.~\ref{sec:gen}, because $I_\nu(0)=0$.
In each case, the effective Wronskian (\ref{effw}) is the same,
\be
\alpha^{E,H}=\frac\alpha2.
\ee
Using the UAE, the leading bulk stress tensor component is
\begin{subequations}\label{bulkexp}
\bea
t^{b,E}_{zz}&=&-\frac{\breve\kappa}2,\quad \breve\kappa=\sqrt{k^2+\zeta^2
e^{\alpha z}},\label{bulkexpa}\\
t^{b,H}_{zz}&=&-\frac{\breve\kappa^2}{2\hat\kappa},\quad 
\hat\kappa=\sqrt{k^2+\zeta^2e^{\alpha z}+\frac{\alpha^2}4}.
\eea
\end{subequations}
The scattering part of the reduced
stress tensor, near the wall, has the form seen 
before in Eq.~(\ref{tsehmod1}) if we replace $\zeta^2$ by 
$\kappa^2\cos^2\theta$, and $a$ by $2/\alpha$, where $\alpha$ 
is the slope of the potential at the edge:
\be
t^{s,E,H}_{zz}\sim-\frac{\alpha^2}{16\kappa}e^{-2\kappa z}
\left\{\begin{array}{c}
\left(\frac{\zeta^2}{2\kappa^2}\right)^2,\\
\left(1-\frac{\zeta^2}{2\kappa^2}\right)^2.\end{array}\right.\label{exptzz}
\ee
From this follows the same result for the stress tensor as in 
Eq.~(\ref{edgeex}). 

We will see in the following section that this behavior is universal, as long
as the potential is continuous and has a linear slope at the edge.

\section{Universal edge behavior}
\label{universal}
\subsection{First-order perturbation theory}
\label{Sec:first-order}
Griniasty and Leonhardt \cite{Griniasty:2017iix} asserted that the behavior 
of the $zz$ component of the subtracted stress tensor seen in 
Eq.~(\ref{edgeex}) is 
universal.  That is, it holds whenever the potential is continuous, but has a
discontinuous slope at the origin, the slope being in that case $\alpha=2/a$.
We will prove that assertion here, which follows from perturbation theory near
the edge.  We can generalize this slightly, by allowing for a 
discontinuity $\epsilon-1$ in the permittivity near the boundary.
Sufficiently close to the edge, $\varepsilon(z)=\epsilon(1+\alpha z)$, and we 
will calculate the stress tensor in the approximation that $\alpha$ is very 
small compared to $\kappa$.  

We start with the TE mode.  The functions $F$ and $G$ satisfy
\be
\left(-\frac{d^2}{dz^2}+\tilde\kappa^2+\zeta^2\epsilon\alpha z\right)
\left\{\begin{array}{c}
                                                            F^E(z)\\G^E(z)
                                                            \end{array}\right.
=0.\label{linpot}
\ee
This is easily solved perturbatively for solutions that decay exponentially
fast, or that grow exponentially fast, at infinity:
\be
\left.\begin{array}{c}
       F^E(z)\\G^E(z)
      \end{array}\right\}=e^{\mp\tilde\kappa z}f^E_{\mp}(z),\quad
f^E_{\mp}(z)=\left(1-\frac{\zeta^2\alpha\epsilon z}
{4\tilde\kappa^2}      (1\pm\tilde\kappa z)\right),\label{firstte}
\ee
keeping terms out through $O(\alpha)$.  Since the differential equation
contains no first derivatives, the Wronskian remains constant,
\be
w(z)=2\tilde\kappa.
\ee
Using the ``bulk'' part of the Green's function in the medium, 
the first term in
the second line of Eq.~(\ref{hegf}), we find for the corresponding 
reduced stress tensor
\be
t_{zz}^{b,E}=-\frac{\tilde\kappa}2
-\frac{\alpha \zeta^2\epsilon}{4\tilde\kappa}z +O(\alpha^2),\label{rbulktzz}
\ee
which agrees with Eqs.~(\ref{tbHEzz}) or (\ref{bulkexpa})
when they are expanded for small $\alpha$ (fixed
$z$).   Integrated over frequency and wavenumbers, we obtain
the full bulk stress tensor, when time-splitting, or 
transverse space-splitting, regulation as in Eq.~(\ref{tsbulk}) is inserted
(see Appendix \ref{appa}),
\begin{subequations}\label{tbulke}
\bea
T_{zz}^{b,E,\tau}&=&\frac1{2\pi^2\sqrt{\epsilon}\delta^4}\left(1+\frac{3}2
\alpha z\right),\quad \bm{\Delta}=0, \delta=\tau/\sqrt{\epsilon},\label{tbezz}
\\
T^{b,E,\delta}_{zz}&=&\frac1{2\pi^2\sqrt{\epsilon}\delta^4}\left(1-\frac{1}2
\alpha z\right),\quad \tau=0,\delta=|\bm{\Delta}|.\label{tbezzb}
\eea
\end{subequations}
The relative factor of $-3$ between the linear dependencies of these two 
forms is the result of the identity given in Ref.~\cite{Milton:2014psa}, 
reproduced here in Eq.~(\ref{identity}).

The reflection coefficient computed from Eq.~(\ref{intrc}) to first order in 
$\alpha$ is
\be
\tilde R^E=-\left(\frac{\kappa-\tilde\kappa}{\kappa+\tilde\kappa}+
\frac{\zeta^2\alpha\epsilon}{4\tilde\kappa^2}\frac1{\kappa+\tilde\kappa}
\right).\label{rtildee}
\ee
The first term in the parentheses refers to the scattering due to the
discontinuity in $\varepsilon(z)$ at the edge, while the second term 
refers to the contribution arising from the slope of the
potential.  If the latter effect
is negligible, this agrees with the form in Eq.~(\ref{rcdiscepsilon}).
A bit of algebra shows the ``scattering'' part of the 
reduced stress tensor is
\be
t_{zz}^{s,E}=-\frac{\alpha\epsilon\zeta^2}{8\tilde\kappa^2}
\left(\frac{\kappa-\tilde\kappa}{\kappa+\tilde \kappa}
+\frac{\zeta^2\alpha\epsilon}{4\tilde\kappa^2}\frac1{\kappa+\tilde\kappa}
\right)e^{-2\tilde\kappa z}.\label{tzzsE}
\ee
When Eq.~(\ref{tzzsE}) is integrated over frequency and 
transverse wavenumbers according to Eq.~(\ref{polarcoords}), 
the result for the stress coincides with that given
in Eq.~(\ref{theepsdisc}):
\be
T^{s,E}_{zz}=-\frac{\alpha E(\epsilon)}{128\pi^2\sqrt{\epsilon}}\frac1{z^3},
\ee
recalling that there $\alpha=2/a$.  On the other hand, if $\varepsilon(z)$
is continuous, so $\epsilon=1$, 
we obtain
\be
T_{zz}^{s,E}=-\frac{\alpha^2}{2560\pi^2}\frac1{z^2}.\label{genedgete}
\ee
This exactly coincides with Eq.~(\ref{edgeex}).

The linearized version of the TM equation (\ref{fgdeq}) is
\be
\left[-\frac{d^2}{dz^2}+\tilde\kappa^2+\alpha\left(\frac{d}{dz}z\frac{d}{dz}
-k^2 z\right)\right]\left\{\begin{array}{c}
                      F^H(z)\\G^H(z)
                     \end{array}\right.=0,
                     \ee
which has a first-order perturbative solution
\be
\left.\begin{array}{c}
       F^H(z)\\G^H(z)
      \end{array}\right\}=e^{\mp\tilde\kappa z}f^H_\mp(z),\quad
f^H_{\mp}(z)=\left[
      1+\frac{\alpha z}2\left(1-\frac{\zeta^2\epsilon}{2\tilde\kappa^2}
(1\pm\tilde\kappa z)\right)\right].
      \ee
Now the Wronskian of the two solutions is not constant,
\be
w^H(z)=2\tilde\kappa(1+\alpha z),
\ee
which is exactly what is needed to make $\alpha^H$ a constant:
\be
\alpha^H= \frac{w^H(z)}{\varepsilon(z)}=\frac{2\tilde\kappa}\epsilon.
\ee
The bulk term in the $zz$-component of the stress tensor turns out to be
the same as its TE counterpart (\ref{tbulke}), for example, for time splitting:
\be
T^{b,H,\tau}_{zz}
=\frac1{2\pi^2\sqrt{\epsilon}\delta^4}\left(1+\frac{3}2\alpha z
\right).\label{tbhzz}
\ee
Now it is straightforward to calculate the 
scattering part of the stress tensor to $O(\alpha^2)$, 
in terms of the reflection coefficient:
\be
\tilde R^H=-\left[\frac{\kappa-\tilde\kappa/\epsilon}{\kappa
+\tilde\kappa/\epsilon}-\frac\alpha{2(\kappa+\tilde\kappa/\epsilon)}
\left(1-\frac{\zeta^2\epsilon}{2\tilde\kappa^2}\right)\right].\label{rh1order}
\ee
The $zz$ component of the scattering part of the reduced stress tensor is then
\be
t_{zz}^{s,H}=\frac{\alpha}{4}e^{-2\tilde\kappa z}\left[\frac{\kappa-\tilde
\kappa/\epsilon}{\kappa+\tilde\kappa/\epsilon}-\frac\alpha{2}
\frac1{\kappa+\tilde\kappa/\epsilon}\left(1-\frac{\zeta^2\epsilon}{2\tilde
\kappa^2}\right)\right] \left(1-\frac{\zeta^2\epsilon}{2\kappa^2}\right).
\label{thzz}
\ee
Again, if for $\epsilon\ne1$ we drop the second term in the square brackets,
we see the appearance of the TM reflection coefficient for a discontinuity
in the permittivity, which leads to the stress tensor as $z\to 0+$:
\be
T^{s,H}_{zz}=-\frac{\alpha}{128\pi^2\sqrt{\epsilon}}H(\epsilon)\frac1{z^3},
\ee
coinciding with the TM part of Eq.~(\ref{theepsdisc}).
If $\epsilon=1$ however, the second term in Eq.~(\ref{thzz}) must be retained,
leaving just the form seen in Eq.~(\ref{exptzz}), and we obtain for the
$zz$ component of the stress tensor
\be
T_{zz}^{s,H}=-\frac{43}{7680\pi^2}\frac{\alpha^2}{z^2},
\ee
which again exactly coincides with Eq.~(\ref{edgeex}).

\subsection{Dispersion}
\label{Sec:disp}
The above assumes that the permittivity does not depend on frequency.  This
is quite unrealistic.  Instead, let's examine what happens if we use a 
plasma model, where $\alpha=\alpha_0/\zeta^2$.  This then makes the TE mode
coincide with the linear scalar problem considered in 
Ref.~\cite{Milton:2016sev}.  There the divergent terms were isolated using
a WKB approximation.  We can easily reproduce those leading divergences.
To compare with the results there, we set the discontinuity $\epsilon-1$ equal
to zero.

With the plasma dispersion relation, the bulk term (\ref{rbulktzz}) 
reads, before integration,
\be
t^{b,E}_{zz}=-\frac\kappa2-\frac{\alpha_0 z}{4\kappa},\label{tezzb}
\ee
and then carrying out the frequency and wavenumber integrations using the
formulas in Appendix \ref{appa}, we find
\be
T^{b,E}_{zz}=\frac1{2\pi^2\delta^4}-\frac{\alpha_0z}{8\pi^2\delta^2},
\label{Tezzb}
\ee
which are the two leading divergent terms found in Ref.~\cite{Milton:2016sev}
for a linear potential.  Perhaps surprisingly, 
the same holds for $T^{b,H}_{zz}$.

To get the logarithmically divergent term in $T^{b,E}_{zz}$ one might think
we would have to work out perturbation theory to second order, which we
will do in the next Section.  However, the $zz$ component of
the reduced bulk stress tensor to second order can be calculated
by knowing only the $O(\alpha)$ solutions because we easily see from the
definition of the Wronskian that
\be
t^{b,E}_{zz}=-\frac{\tilde\kappa}2+\frac1{2w^E}(f^{E\prime}_-f^{E\prime}_+
-\alpha \zeta^2\epsilon z f^E_-f^E_+).\label{bulkstred}
\ee
  From this follows
\be
t^{b,E}_{zz}=-\frac{\tilde\kappa}2-\alpha\gamma\tilde\kappa z
+\alpha^2\gamma^2\left(
\frac1{4\tilde\kappa}+\tilde\kappa z^2\right),\label{tzzbpert}
\ee
where we have introduced the abbreviation 
$\gamma=\zeta^2\epsilon/(4\tilde\kappa^2)$.
%
The small $\delta$ expansion  of $\int_0^\infty d\kappa \sin(\kappa
\delta)/\kappa^2$ [Eq.~(\ref{a3e})] yields in second order in the plasma
model
\be
T_{zz}^{b,E (2)}\sim -\frac{\alpha_0^2 z^2}{32\pi^2}\ln\delta,\label{logdive}
\ee
which corresponds
to the logarithmically divergent term found in Ref.~\cite{Milton:2016sev}.

Again only the first order result is necessary to give the order
$\alpha^2$ contribution to the bulk stress for the TM mode, 
since the same formula as 
Eq.~(\ref{bulkstred}) applies for the TM mode as well.  The result is only
slightly different from that in Eq.~(\ref{tzzbpert}):
\be
t^{b,H}_{zz}=-\frac{\tilde\kappa}2-\alpha\gamma\tilde\kappa z+
\alpha^2\gamma^2\tilde\kappa z^2
+\frac{\alpha^2}{4\tilde\kappa}\left(\gamma-\frac12\right)^2.
\ee
This leads to exactly the same logarithmic divergence in the plasma model 
as in Eq.~(\ref{logdive}).
However, to get such terms for the other components of the stress tensor,
we need second-order perturbative solutions for $F$ and $G$, which
we will deal with in the following section.

As for the scattering contributions, it is evident that due to the softening
produced by the plasma dispersion relation, the singular behavior in 
$T^{s,E}_{zz}$
as the edge is approached from the inside goes away, consistent with the
numerical results shown in Fig.~5 of Ref.~\cite{Milton:2016sev}.  (See
further discussion in Sec.~\ref{sec:exlin}.)
The scattering part of $T^{s,H}_{zz}$ in the plasma model has to be
defined with an infrared cutoff, but certainly 
also does not diverge as the edge is approached. 

So we have verified and extended the results of 
Ref.~\cite{Griniasty:2017iix}:  For a vacuum interface with a planar
dielectric without dispersion,  if the permittivity is continuous,
but has a linear slope at the edge, the singularities in the normal-normal
 component of the stress tensor possess  a universal $1/z^2$ form,
where $z$ is the distance from the edge into the medium.  
If the permittivity is discontinuous,
the normal-normal component of the  stress tensor has a $1/z^3$
singularity, and as shown in Appendix \ref{appb},
the singularity is reduced to logarithmic if the discontinuity is in the
second derivative.  As we will see in the next Section, 
the singularities in the energy density are one order higher for a linear
discontinuity.   Only the behavior
of the potential at the edge of the dielectric is necessary to determine
the singularities in form and magnitude, but this we have demonstrated
through examples and a general perturbative analysis. 

\section{Other stress tensor components}
\label{other}
Let us now examine other components of the stress tensor, particularly in the 
continuous permittivity situation.  The leading perturbative approximation
yields  the leading divergent structure, and the leading
 behavior near the edge.  We will  consider both
the dispersive case with the plasma model, since that agrees,
for the TE mode, with the scalar case, and is approximately realistic,
as well as the situation when the permittivity is independent of frequency.

\subsection{Leading-order contributions}
Including the dispersive factor, the reduced TE energy density for the plasma 
model, where $\alpha=\alpha_0/\zeta^2$, is for small $\alpha_0z$ (exactly, 
for a linear potential)
\be
t^E_{00}=
\frac12\left(\partial_z\partial_{z'}+k^2-\zeta^2\epsilon+\alpha_0 \epsilon
z\right)g^E,\label{te00}
\ee
which agrees with the scalar energy density for a linear potential provided
the conformal parameter $\xi=0$ (or in the language of 
Ref.~\cite{Milton:2016sev}, 
$\beta=-1/4$), surprisingly, not the scalar conformal value of $\xi=1/6$.
(That is, the canonical stress tensor emerges, not the conformal one.) 
Thus we see that (setting $\epsilon=1$)
\be
t^E_{00}=t^E_{zz}+(k^2+\alpha_0z)g^E.
\ee
Using the point-splitting methods of the Appendix, we find for
the bulk contribution to the energy density,
\be
T_{00}^{b,E}\sim\left\{\begin{array}{cc}
\dfrac3{2\pi^2\delta^4}-\dfrac{\alpha_0z}{8\pi^2\delta^2},
&\tau\,{\rm splitting},\\
\noalign{\smallskip}\\
-\dfrac1{2\pi^2\delta^4}+\dfrac{\alpha_0z}{8\pi^2\delta^2},
&\Delta\,{\rm splitting},\end{array}\right.\label{te00b}
\ee
which coincides with the leading divergences found in
Ref.~\cite{Milton:2016sev}.  Note that 
\be
\frac\partial{\partial\delta}\left(\delta T_{00}^\Delta\right)=T_{00}^\tau
\ee
holds for the relation between the energy densities with the spatial and 
temporal cutoffs, as in Ref.~\cite{Milton:2014psa}.
And in the medium, just to the right of the edge, we find for the
scattering contribution
\be
t^{s,E}_{00}\sim-\frac{\alpha_0 k^2}{16\kappa^4}e^{-2\kappa z},
\ee
which when integrated over frequency and wavenumbers yields
\be
T_{00}^{s,E}\sim-\frac{\alpha_0}{96\pi^2z}, \label{t00z}
\ee
 exactly the result as for the scalar case with $\beta=-1/4$ given by
Eq.~(6.7) of Ref.~\cite{Milton:2016sev}.

Had we assumed that $\alpha$ was independent of $\zeta$, 
 the sign of the potential term in Eq.~(\ref{te00}) 
would have reversed, and we would have
obtained instead  for the bulk divergence (with temporal splitting)
\be
T^{b,E}_{00}=\frac3{2\pi^2\delta^4}\left(1+\frac32\alpha z\right),
\label{tbe00}
\ee
and for the edge singularity in the scattering part
\be
T^{s,E}_{00}\sim -\frac\alpha{960\pi^2 z^3},\label{tse00}
\ee
more singular than the behavior of $T^{s,E}_{zz}$ in this non-dispersive
model seen in Eq.~(\ref{genedgete}).

For the remaining diagonal components, from Eq.~(\ref{txxtyy}),
for $\tau$ splitting, or $\bm{\Delta}$ splitting, respectively, in the
plasma model,
\be
T^{b,E}_{xx}=T^{b,E}_{yy}=\left\{\begin{array}{c}
\dfrac1{2\pi^2\delta^4}-\dfrac{\alpha_0 z}{8\pi^2
\delta^2},\\
\noalign{\smallskip}\\
-\dfrac1{2\pi^2\delta^4},\end{array}\right.\label{Texxb}
\ee
which exactly coincides with the leading scalar divergences found in 
Ref.~\cite{Milton:2016sev} when we average over $\rho_x$, $\rho_y$ there.  
It is easily checked that the trace identity
(\ref{traceid}) is satisfied:
\be
(T^{b,E})^\mu{}_\mu=-\frac{\alpha_0 z}{4\pi^2\delta^2}.
\ee
For the scattering part,
\be
T^{s,E}_{xx}=T^{s,E}_{yy}=-\frac{\alpha_0}{192\pi^2z},\label{tsExx}
\ee
which is exactly half the energy density found in Eq.~(\ref{t00z}) as
required by the trace of the scattering part of the
stress tensor being of $O(\alpha_0^2)$.

For the  TM mode in the plasma model 
the bulk part of the reduced energy density is
\be
t_{00}^{b,H}=-\frac{\zeta^2}{2\kappa}\left(1-\frac{\alpha_0z}{2\kappa^2}
\right),
\ee
which, upon integration, leads to the same result as Eq.~(\ref{te00b}).
The transverse bulk parts of the reduced stress tensor are
\be
t^{b,H}_{xx}=t^{b,H}_{yy}=\frac{k^2}{4\kappa}\left(1-\frac{\alpha_0 z}{2
\kappa^2}\right), 
\ee
 leading to the same result as Eq.~(\ref{Texxb}), as required by the
trace identity.   For constant $\alpha$ the energy density
divergence is the same  as for the TE part, Eq.~(\ref{tbe00}).
The scattering part of the reduced energy density is
\be
t^{s,H}_{00}=\frac{\alpha_0}{8\zeta^2}\frac{k^2}{\kappa^2}\left(1
-\frac{\zeta^2}{2\kappa^2}\right)e^{-2\kappa z}=2 t^{s,H}_{xx}=2t^{s,H}_{yy},
\label{t00a}
\ee
which is twice  the transverse reduced
stress tensor components, as required by the trace identity.
This possesses singularities, 
when $\zeta^2=\kappa^2\cos^2\theta$ goes to
zero, so the meaning of these  seems somewhat obscure.
However, if we  adopt the nondispersive model
and assume that $\alpha$ is
constant,  we can find the energy density singularity near the edge
\be
T_{00}^{s,H}=\frac{3\alpha}{320\pi^2 z^3},
\ee
which is $-9$ times that from the TE mode, Eq.~(\ref{tse00}).
\subsection{Second order perturbation theory}
To proceed further, we need to work to the next order in perturbation theory.
It is easy to work out the solutions to Eq.~(\ref{linpot}) to second order,
assuming the potential is exactly linear.  The two solutions are
\be
\left.\begin{array}{c}
F^E(z)\\G^E(z)
\end{array}\right\}=
e^{\mp\tilde\kappa z}\left(1-\alpha\gamma z(1\pm\tilde\kappa z)
+\frac{\alpha^2\gamma^2z}
{\tilde\kappa}\left[\frac12(\tilde\kappa z)^3\pm\frac53(\tilde\kappa z)^2+
\frac52\tilde\kappa z\pm\frac52\right]\right)+O(\alpha^3)\equiv
e^{\mp\tilde\kappa z}f^E_{\mp}.\label{2ndfe}
\ee
The expansion parameter is $\alpha\gamma$.
The Wronskian changes, but is still  constant,
\be
w^E=2\tilde\kappa-5\frac{\alpha^2\gamma^2}{\tilde\kappa}+O(\alpha^3).
\ee

The TM equation (\ref{fgdeq}) is, assuming an exactly linear potential,
\be
\left(-\frac\partial{\partial z}\frac1{1+\alpha z}\frac\partial{\partial z}
+\frac{k^2}{1+\alpha z}+\zeta^2\epsilon\right)\left\{
\begin{array}{c}F^H\\G^H\end{array}\right.=0,
\ee
which can also be straightforwardly solved to second order in $\alpha$:
\bea
\left.\begin{array}{cc}
F^H(z)\\G^H(z)\end{array}\right\}&=&e^{\mp\tilde \kappa z}\bigg\{1 +z\left[
\alpha\left(\frac12-\gamma\right)\mp\frac{\alpha^2}{2\tilde\kappa}
\left(\frac{3}4
-5\gamma^2\right)\right]+z^2\left[\mp\tilde\kappa\alpha\gamma+
\alpha^2\left(\frac{5\gamma^2}2-\frac{1}8
-\frac{\gamma}2\right)\right]\nn\\
&&\quad\mbox{}\mp\frac{\alpha^2\gamma\tilde\kappa z^3}6(3-10\gamma)
+\frac{\alpha^2\gamma^2\tilde\kappa^2 z^4}2\bigg\}.\label{2ndtm}
\eea
Note that the terms of order $\alpha\gamma$ and of order $\alpha^2\gamma^2$
coincide with those of the TE solutions in Eq.~(\ref{2ndfe}).
The Wronskian of these two solutions gives
\be
\alpha^H=\frac{w^H}{\epsilon(1+\alpha z)}=\frac{2\tilde\kappa}\epsilon+
\alpha^2\frac{3-20\gamma^2}{4\tilde\kappa\epsilon}.
\ee

\subsection{$O(\alpha^2$) corrections}
Now to get the order-$\alpha^2$ corrections
to the energy density, we have to use the second-order
solutions, Eqs.~(\ref{2ndfe}) and (\ref{2ndtm}).
A straightforward calculation reveals, for the bulk contributions to the
reduced energy density,
\begin{subequations}
\be t_{00}^{b,E}=-2\tilde\kappa\gamma+4\alpha \tilde\kappa\gamma^2z
+\frac{\alpha^2\gamma^2}{2\tilde\kappa}(3-10\gamma-24\gamma\tilde
\kappa^2z^2),
\ee
and
\be
t_{00}^{b,H}=-2\tilde\kappa\gamma+4\alpha\tilde\kappa\gamma^2 z
+\frac{\alpha^2}{8\tilde\kappa}(-1+4\gamma+12\gamma^2-40\gamma^3
-96\gamma^3\tilde\kappa^2z^2).
\ee
\end{subequations}  
Note that the
$O(\alpha^0)$, $O(\alpha)$ and the $O(\alpha^2\gamma^2)$, $O(\alpha^2\gamma^3)$
 terms are the same for the TE and TM contributions, which means that the
divergences in the plasma model are the same, for example, in temporal point
splitting  for $\epsilon=1$ as defined in Appendix \ref{appa}, 
\be
T_{00}^{b,E,H}=\int\frac{d\zeta}{2\pi}\int\frac{(d\mathbf{k}_\perp)}{(2\pi)^2}
e^{i\zeta\tau}
t_{00}^{b,E,H}=\frac{3}{2\pi^2\delta^4}-\frac{\alpha_0 z}{8\pi^2}\frac1
{\delta^2}+\frac{\alpha_0^2z^2}{32\pi^2}\ln\delta+\dots,
\ee
where the remainder is finite as $\delta\to0$.  This includes the results
already found in Eq.~(\ref{te00b}), and coincides with the scalar divergences 
found in Ref.~\cite{Milton:2016sev}.  

We can also straightforwardly find the next order corrections to the scattering
part of the $zz$ component of the reduced TE stress tensor, 
for example, with $\epsilon=1$,
\be
t_{zz}^{s,E}=-\frac{\alpha^2\gamma^2}{4\kappa}[1-2\alpha\gamma z(2+\kappa z)]
e^{-2\kappa z},
\ee
 but the order
$\alpha^3$ correction means that the corresponding term in $T_{zz}^{s,E}$ has
one less power of $z$, so in the constant $\alpha$ situation, through this 
order,
\be
T_{zz}^{s,E} 
=\frac{\alpha^2}{2560\pi^2z^2}+\frac{3\alpha^3}{1768\pi^2z}.
\ee
(In the plasma model, recall that there is no singularity in $T^{s,E}_{zz}$.)
Dimensionally, since $[\alpha] 
=1/L$, the higher order corrections to the edge singularity must be 
subdominant.  

Similarly, we can write for the TE part of the reduced energy density 
through order $\alpha^2$,
\be
t_{00}^{s,E}=-\alpha\gamma\frac{k^2}{4\kappa^2}\left(1-\frac{\alpha\gamma}
{\kappa}\left[k^2(-2+2\kappa z+2(\kappa z)^2)-\kappa^2(1+4\kappa z)\right]
\right)e^{-2\kappa z},
\ee
which leads to, for constant $\alpha$, the energy density through $O(\alpha^2)$
\be
T_{00}^{s,E}=-\frac\alpha{960\pi^2 z^3}-\frac{\alpha^2}{17920\pi^2z^2},\quad
z\to 0+.\ee
Again, the correction is necessarily subdominant.

\section{Exact linear TE potential}
\label{sec:exlin}
Of course, the linear TE problem is exactly solvable in terms of Airy 
functions, as seen in Refs.~\cite{Bouas:2011ik,Murray:2015tim,
Milton:2016sev,Milton:2011iy}.
 Independent solutions of Eq.~(\ref{linpot}) are ($\alpha=\alpha_0/\zeta^2$)
\be
\left.\begin{array}{c}F(z)\\G(z)\end{array}\right\}=\left\{\begin{array}{c}
\Ai\bigg(\alpha_0^{-2/3}(\kappa^2+\alpha_0z)\bigg),\\ 
\Bi\bigg(\alpha_0^{-2/3}(\kappa^2+\alpha_0z)\bigg),\end{array}\right.
\ee
which have Wronskian $\alpha_0^{1/3}/\pi$.  It is then immediate to write
down the exact form of the Green's function.

\begin{figure}
\includegraphics{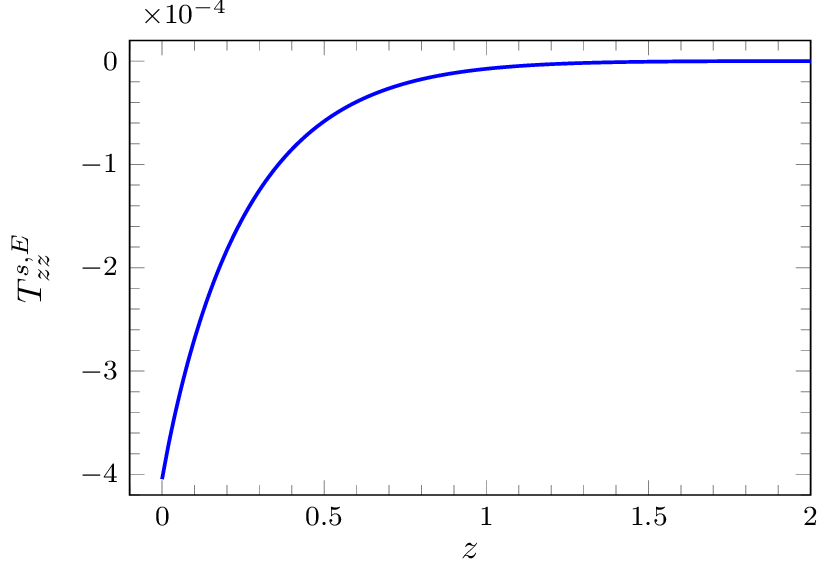}
\caption{\label{figtzz} The exact TE scattering contribution
to the $zz$ component of the stress tensor $T^{s,E}_{zz}$
within a medium having a linear potential, $\varepsilon(z)=1+z$,
characterized by a plasma-model dispersion relation.
(That is, $\epsilon=1=\alpha_0$.)  Although
the stress gets larger in magnitude as the edge is approached, it remains
finite, and it goes to zero deep within the medium.}
\end{figure}

By using the asymptotic expansion of the Airy functions for large argument, 
we straightforwardly obtain for the TE reduced scattering Green's function
\be
g^{s,E}(z,z')\sim -\frac{\alpha_0}{16\kappa^3}
\frac{\exp\left[\frac{2\kappa^3}{3
\alpha_0}\left(2-(1+\alpha_0 z/\kappa^2)^{3/2}
-(1+\alpha_0z'/\kappa^2)^{3/2}\right)\right]}
{[(\kappa^2+\alpha_0z)(\kappa^2+\alpha_0 z')]^{1/4}}.  
\ee  The above is valid if $\kappa^3/\alpha_0\gg1$.  If we now regard the
potential as weak, we expand in powers of $\alpha_0$ and obtain through
second order
\be
g^{s,E}(z,z')\approx -\frac{\alpha_0}{16\kappa^4}\left(1-\frac{\alpha_0}{4
\kappa^2}(z+z')-\frac{\alpha_0}{4\kappa}(z^2+z^{\prime2})\right)
e^{-\kappa(z+z')}.
\ee
This coincides exactly with the Green's function
 obtained from the perturbative solution
(\ref{firstte}),  and leads, for example, to
\be
t^{s,E}_{zz}=-\frac{\alpha_0^2}{64\kappa^5}e^{-2\kappa z},
\ee
which follows from (\ref{tzzsE}) 
when $\epsilon=1$ and $\alpha=\alpha_0/\zeta^2$.
But when one tries to integrate this over wavenumbers and frequency, one
encounters an infrared divergence at $\kappa=0$.  Of course, such a divergence
is not present in the exact solution, since the perturbative expansion
is not valid for small $\kappa$.
In fact, if the exact expression for $t^{s,E}_{zz}$ is integrated the
result is finite, but nonzero, at $z=0$, as shown in Fig.~\ref{figtzz},
as earlier stated.

\begin{figure}
\includegraphics{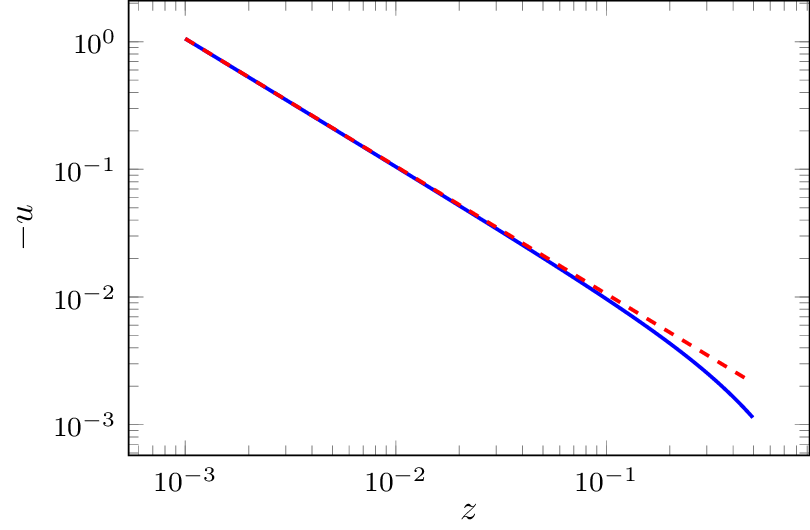}
\caption{\label{t00fig} The TE scattering contribution to the energy
density within the medium, again calculated in the plasma model, with
$\varepsilon(z)=1+z$.  The solid
curve is the exact numerical integration, which has to be carried out to
very large values of $\kappa$ for small $z$, because of near-perfect
cancellations between the moderate $\kappa$ contributions.  The dashed
curve represents the asymptotic estimate (\ref{t00z}).}
\end{figure}

We can do the same type of calculation for the energy density.  In this
case the energy density does diverge as the edge is approached from within
the medium, according to Eq.~(\ref{t00z}).  In fact, the numerical integration
of the exact formula fits this asymptotic formula quite well for small $z$, 
as shown in Fig.~\ref{t00fig}.
$T^{s,E}_{xx}$ has nearly identical behavior, except for the factor of 2 seen
in Eq.~(\ref{tsExx}).

\begin{figure}
\includegraphics{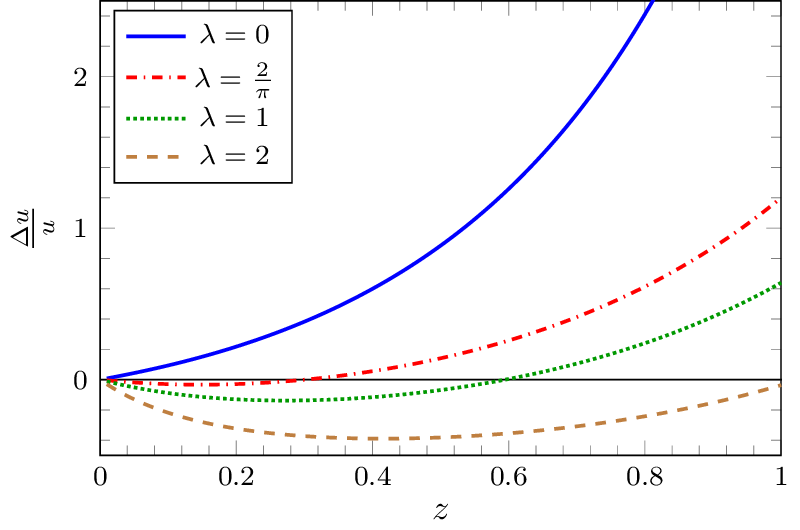}
\caption{\label{figlambda} The relative error of the asymptotic estimate
for the TE scattering
energy density (\ref{t00z}) when the reflection coefficient $\tilde
R^E$ is replaced by $\tilde R^E-\lambda\pi/2$.  Here $u=T^{s,E}_{00}$
and $\Delta u=(T^{s,E}_{00})_{\rm asym}-T^{s,E}_{00}$.
Shown are the errors for $\lambda
=0$ (upper curve), that is, just using the $\Bi$ function as the second 
solution, and for $\lambda = 2/\pi, 1, 2$, that is, with $\Bi$ 
replaced by different mixtures of $\Bi$ and $\Ai$. 
Here again we assume $\alpha_0=1$.}
\end{figure}

The above figures were drawn with the assumption that the second solution $G$
was exactly the second Airy function $\Bi$.  But, as noted in 
Sec.~\ref{sec:gen}, the 
definition of the reflection coefficient is ambiguous, since the second
solution may contain an arbitrary admixture of the first.  
 The criteria given in Sec.~\ref{sec:gen} do
not apply, because both $\Ai$ and $\Bi$ behave as damped oscillatory functions
for large negative $z$. The addition of the second solution is typically
asymptotically exponentially subdominant, so this ambiguity does not
appear in the asymptotic estimates.  However, the ambiguity will affect the
behavior away from the edge. 
 We investigated this by substituting in the
reflection coefficient $\Bi\to\Bi+\lambda \Ai$, where $\lambda$ is a constant.
(In fact, $\lambda$ could be a function of $\kappa$.)
In Fig.~\ref{figlambda} we show how agreement with the estimate
(\ref{t00z}) is greatly improved by the choice of $\lambda=2/\pi$.  The reason
for this particular value agreeing with the asymptotic
estimate is, at present, mysterious.

The edge singularity is not altered when different constant
values of $\lambda$ are used as compared to the perturbative result
 because the leading asymptotic behavior of
 the Airy functions is
\be
\left.\begin{array}{c}\Ai(x)\\ \Bi(x)\end{array}
\right\}=\frac1{(2,1)\sqrt{\pi}x^{1/4}}e^{\mp 2 x^{3/2}/3},\quad x\to+\infty,
\ee
so that when these are used for large $\kappa$ and fixed $z$ we see that the
admixture parameter is related to the perturbation theory one by
\be
\lambda_{\rm PT}=\frac\lambda2 e^{-4\kappa^3/3}.
\ee
Here, the latter parameter is defined in the language of 
Sec.~\ref{Sec:first-order} by taking the second solution to be
\be
G=e^{\kappa z}f_++\lambda_{\rm PT}e^{-\kappa z}f_-.
\ee
Thus, it is evident that the admixture of the first solution will be 
exponentially suppressed within the wavenumber integral.
 
\begin{figure}
\includegraphics{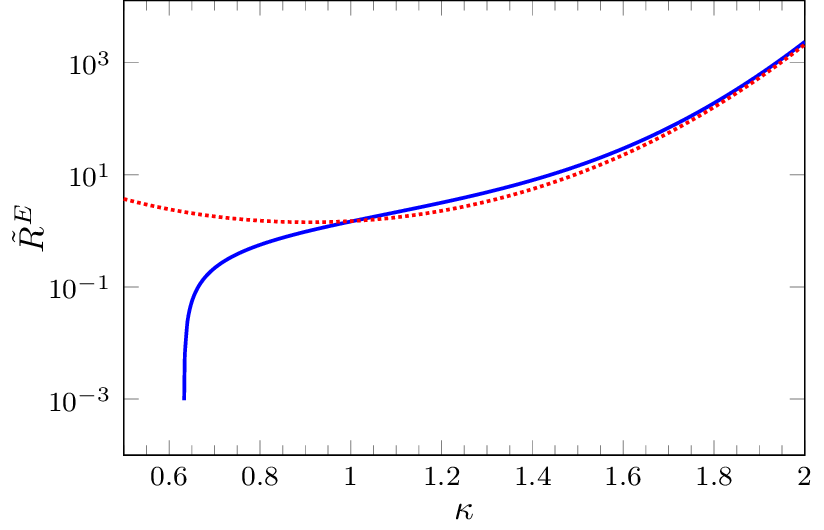}
\caption{\label{fig:rc} The asymptotic TE  reflection coefficient 
$-\alpha_0/(8\kappa^3)$,
Eq.~(\ref{rtildee}), (dotted) compared
to the exact reflection coefficient (solid) given by Eq.~(\ref{intrc}),
for the linear potential.  The former
has to be normalized by the correct factor to account for the normalization of
the Airy functions in the Green's function.  Here $\alpha_0=\epsilon=1$.}
\end{figure}

The comparison between the perturbative value of the reflection coefficient
and the exact one is shown in Fig.~\ref{fig:rc}.
Because the perturbative solutions are normalized such that $F(0)=G(0)=1$,
which is not the case for the Airy functions,  an appropriate normalization
factor must be supplied: What is plotted in the dotted curve in the
figure is $R_{\rm PT}=-\frac{\pi}{8\kappa^3} e^{4\kappa^3/3}$.  These curves
reveal that the validity of the perturbative solution depends on the inequality
\be
\alpha_0\ll \kappa^3.
\ee

\begin{figure}
\includegraphics{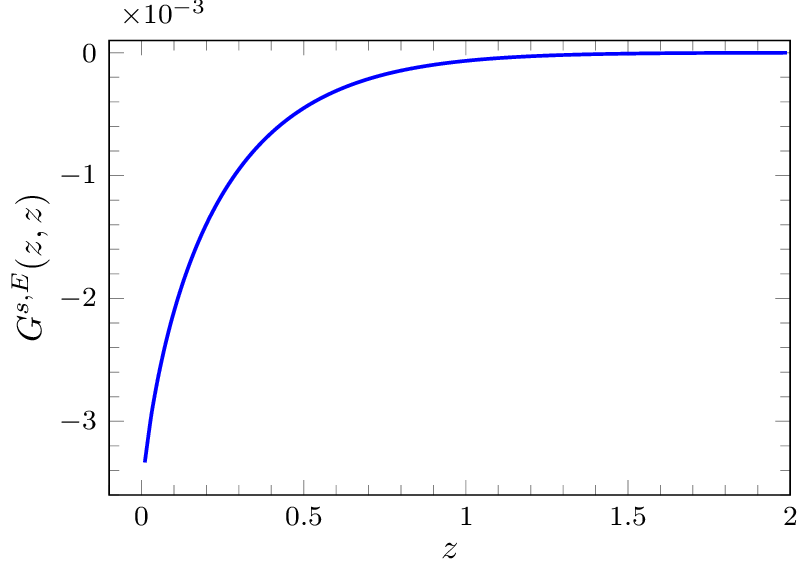}
\caption{\label{fig:gzz} The diagonal elements of the 
scattering contribution to the TE Green's function
for the linear wall, $\varepsilon(z)=1+z$, for $z$ within the wall.  This
represents the expectation values of the square of the electric field,
which rapidly decrease to zero as the wall is penetrated.}
\end{figure}

It will be noted from Figs.~\ref{figtzz} and \ref{t00fig} 
that the stress tensor components rapidly
go to zero as one goes deeper into the potential, as expected.  To further
explore this, we look at the Green's function, which represents the expectation
value of the product of the electric fields in the medium, for the
case $\alpha_0=1$,
\be
G^{s,E}(z,z')=\int\frac{d\zeta}{2\pi}\frac{(d\mathbf{k}_\perp)}{(2\pi)^2}
g^{s,E}(z,z)=
-\frac1{2\pi}\int_0^\infty d\kappa\,\kappa^2\frac{\kappa\Bi(\kappa^2)
-\Bi'(\kappa^2)}{\kappa\Ai(\kappa^2)-\Ai'(\kappa^2)}\Ai(\kappa^2+z)
\Ai(\kappa^2+z').
\ee
This is plotted, for $z=z'$, in Fig.~\ref{fig:gzz}.
For even larger $z$ than shown in the figure, the reflection coefficient
may be replaced by its small-$\kappa$ expansion, and then the resulting
analytic form of the diagonal Green's function ultimately agrees with that
found by numerical integration.  (For a twenty-term expansion of $\tilde R^E$,
the error of  the analytic approximation is less than 1\% for $z>11$.)

 \section{Reflected potentials}
\label{sec:mirror}
Of course, there is no net force on the semi-infinite slab we have been considering
to this point.  This is because $T_{zz}$ must vanish at infinity, and once
the obvious bulk subtraction is made, $T_{zz}(0-)=0$, according to 
Eq.~(\ref{outstress}).  So suppose we consider two bodies, constructed by
placing the mirror image of our potential to the left of $z=0$: that is, we
assume $\varepsilon(z)=\varepsilon(-z)$. 
 These are two bodies in contact, not disjoint.
Then for either the TE or TM mode, the Green's function may be constructed in
terms of the fundamental solutions of the homogeneous equations, $\tilde F$ 
and $\tilde G$, where
$\tilde F\to0$ as $z\to+\infty$, and $\tilde G\to0$ as $z\to -\infty$,
\be
g(z,z')=\frac1A \tilde F(z_>)\tilde G(z_<).
\ee in terms of the effective Wronskian factor $A$.  If we expand this out in 
terms of the  solutions on the right for the semi-infinite slab,  denoted 
as previously by $F$ and $G$, we find for $z,z'>0$
\be
g(z,z')=\frac1\alpha[F(z_>) G(z_<)+R F(z)F(z')],
\ee
where $\alpha$ is the Wronskian term for the half-space.  Here the reflection
coefficient is
\be
R=-\frac{( F G)'(0)}{(F^2)'(0)}.
\ee

Perturbatively, it is easy to check that to first order
\be
R=\left\{\begin{array}{cc}
-\frac{\alpha\gamma}{\kappa},&\mbox{TE},\\
\frac\alpha\kappa\left(\frac12-\gamma\right),&\mbox{TM},
\end{array}\right.
\ee
which are twice as big as the values found for the semi-infinite slab, in
Eqs.~(\ref{rtildee}) and (\ref{rh1order}), as would be expected, because the 
slope discontinuity is doubled.

In the case of the plasma model, $T^s_{zz}$ is finite, and for an exact linear
potential was solved explicitly in Sec.~\ref{sec:exlin}---see 
Fig.~\ref{figtzz}.
So in the case of two facing  reflected linear potentials
in contact, one might think that
a finite force of one body upon the other could be determined, 
\be
T_{zz}^{s,E}(0)=-0.001017\alpha_0^{4/3},
\ee
where we have restored the proper scaling with the coupling.
Although this appears to be a finite attraction between the two slabs, the 
interpretation  of this is suspect for the reasons stated in 
Sec.~\ref{sec:force}, because the body is not immersed in
a homogeneous medium.
As there is no distance scale in the problem aside from the coupling, it 
is impossible to connect this to a change in the energy according to the 
principle of
virtual work.  Moreover, the ambiguity of separating bulk and scattering parts
remains.

\section{Conclusion}
\label{sec:concl}
In this paper we have extended our previous calculations on the ``soft wall'' 
problem to the electromagnetic case.  In the plasma dispersion model, the TE 
mode coincides
with the scalar case considered in Ref.~\cite{Milton:2016sev}.  Without 
dispersion
we recover the universal edge behavior found by Ref.~\cite{Griniasty:2017iix}.
We also reproduce the Weyl divergences found in the scalar case.  We do this,
first by considering explicitly solvable examples, and then by performing
 a generic
perturbative analysis for small slopes in the dielectric potential.

Let us summarize the salient features.  For the plasma model, where
the potential may be defined by $\varepsilon(z)-1=v(z)/\zeta^2$ we see
universal Weyl singularities in the bulk stress tensor for both TE and
TM polarizations:
\begin{subequations}
\bea
T^{b,E,H}_{zz}&=&
\frac1{2\pi^2\delta^4}-\frac{v}{8\pi^2\delta^2}-\frac{v^2}{32\pi^2}
\ln \delta,\\
T^{b,E,H}_{00}&=&\frac3{2\pi^2\delta^4}-\frac{v}{8\pi^2\delta^2}+\frac{v^2}{32\pi^2}
\ln \delta+\frac{v''}{48\pi^2}\ln\delta,
\eea
\end{subequations}
which coincide with the divergences found for a scalar field 
\cite{Milton:2016sev}.  (The second derivative term is seen for the
quadratic potential treated in Appendix \ref{appb}.)
For the nondispersive model,  with temporal splitting,
\be
T^{b,E,H}_{zz}=\frac1{2\pi^2\delta^4}\left(1+\frac32\alpha z\right),\quad
T^{b,E,H}_{00}=\frac3{2\pi^2\delta^4}\left(1+\frac32\alpha z\right).
\ee
For the singularities just inside the edge, with a constant (non-dispersive)
linear potential near the edge, with no discontinuity,
\begin{subequations}
\bea
T_{zz}^{s,E}&\sim&-\frac{\alpha^2}{2560\pi^2}\frac1{z^2},\quad
 T_{zz}^{s,H}\sim-\frac{43\alpha^2}{7680\pi^2}\frac1{z^2},\\
T_{00}^{s,E}&\sim&-\frac{\alpha}{960\pi^2}\frac1{z^3},\quad
 T_{00}^{s,H}\sim\frac{3\alpha}{320\pi^2}\frac1{z^3}.
\eea
\end{subequations}
These results are very similar to those seen for the quadratic
potential treated in  Appendix \ref{appb}, with the replacements
$\alpha/z\to -\beta$, $\alpha^2/z^2\to(\beta^2/4) \ln z$.

One might think one could remove the Weyl divergences by removing all terms
with polynomial growth in $z$, for surely such growth deep within the material
is unphysical.  Unfortunately, the WKB analysis of Ref.~\cite{Milton:2016sev} 
shows there
must also be $z^2\ln z$ terms in the linear plasma-model TE case, which is
confirmed by numerical experiments, so such a procedure appears impossible.

Although we recover expected results, as well as some new features, our 
analysis
remains incomplete.  It hinges on a break-up between bulk and scattering 
contributions, which is not unique; however, it captures the essential 
asymptotic behavior for large wavenumbers. The suggestion that to 
achieve a finite
stress one merely omits the bulk terms is plausible, but this is not a unique 
process. Moreover, there are finite, position-dependent
contributions to the stress tensor contained in the bulk term that likely
cannot be merely discarded.

\acknowledgments
We thank the US National Science Foundation, grant number 1707511, 
for partial support of this
research.  PP acknowledges the
Research Council of Norway, project number 250346 for support. 
KAM thanks the the Norwegian University of Science
and Technology and the Centro Universitario de la Defensa for their
hospitality during part of the course of this work.
\appendix

\section{Point-splitting regularization}
\label{appa}
To pass from the reduced (Fourier-transformed) stress tensor components to
the space-time stress tensor, we need to integrate over (imaginary) frequency
and transverse wavevectors.  Doing so leads to divergences for the ``bulk''
parts, so we regulate the integrals by point-splitting in the transverse
directions and in time:
\be
T(z)=\int_{-\infty}^\infty \frac{d\zeta}{2\pi}\int\frac{(d\mathbf{k})}
{(2\pi)^2}e^{i\zeta\tau}e^{i\mathbf{k}\cdot\bm{\Delta}}
t(\tilde\kappa,\zeta),\quad \tilde\kappa=\sqrt{k^2+\zeta^2\epsilon},\quad
\tau, \bm{\Delta}\to0,
\ee
writing in a generic form.  If the function $t$ only depends on $\tilde\kappa$
we can evaluate this in polar coordinates, with the polar angle being the
angle  between $\bm{\delta}=(\tau/\sqrt{\epsilon},\bm{\Delta})$ and 
$\bm{\tilde\kappa}=(\sqrt{\epsilon}\zeta,\mathbf{k}).$  Then
\be T(z)=\frac1{2\pi^2}\frac1{\sqrt{\epsilon}}\int_0^\infty d\tilde\kappa\,
\tilde\kappa^2 \frac{\sin\tilde\kappa\delta}{\tilde\kappa\delta}
t(\tilde\kappa).
\ee
The resulting Fresnel integrals of this type are obtained by integrating
by parts and discarding the contribution at infinity 
(justified in a distributional sense):
\begin{subequations}
\bea
\int_0^\infty d\kappa \,\kappa^2\sin\kappa\delta&=&-\frac2{\delta^3},\\
\int_0^\infty d\kappa \,\kappa\cos\kappa\delta&=&-\frac1{\delta^2},\\
\int_0^\infty d\kappa \,\sin\kappa\delta&=&\frac1{\delta}.
\eea
We can also give the integrals which have infrared singularities (regulated by
a cutoff $\mu$, which never appears in the results):
\bea
\int_\mu^\infty\frac{d\kappa}{\kappa}\cos\kappa\delta
&\sim& -\gamma-\ln\mu\delta,\quad \delta\to 0,\\
\int_\mu^\infty\frac{d\kappa}{\kappa^2}\sin\kappa\delta
&\sim&\delta(1 -\gamma-\ln\mu\delta),\quad \delta\to 0.\label{a3e}
\eea
\end{subequations}
But we also encounter terms where $\zeta^2$ appears linearly.  Then it is
easiest to consider time-splitting and space-splitting separately.  For
the $\tau$ cutoff, the angular average of $\epsilon\zeta^2$ gives
\be
T_\tau(z)=\frac1{4\pi^2\sqrt{\epsilon}}\int_0^\infty d\tilde\kappa\,\tilde
\kappa^2\int_{-1}^1 d\cos\theta \,e^{i\tilde\kappa\tau\cos\theta
/\sqrt{\epsilon}}\tilde\kappa^2\cos^2\theta
=-\frac1{2\pi^2\sqrt{\epsilon}}\int_0^\infty d\tilde\kappa\,\tilde\kappa^4
\left(\frac{\partial}{\partial(\tilde\kappa\delta)}\right)^2\frac{\sin\tilde
\kappa\delta}{\tilde\kappa\delta},\quad \delta =\tau/\sqrt{\epsilon},
\ee
while for the spatial cutoff (which, without loss of generality we can
choose to be in the $x$ direction),
\bea
T_\Delta(z)&=&\frac1{8\pi^3\sqrt{\epsilon}}\int_0^\infty d\tilde\kappa\,
\tilde\kappa^2\int_{-1}^1 d\cos\theta\int_0^{2\pi}d\phi\,e^{i\tilde\kappa
\delta\sin\theta\cos\phi}\tilde\kappa^2\cos^2\theta\nn\\
&=&\frac1{4\pi^2\sqrt{\epsilon}}\int_0^\infty d\tilde\kappa\,\tilde\kappa^4
\int_0^\pi d\theta\sin\theta\cos^2\theta J_0(\tilde \kappa\delta\sin\theta)
=\frac1{2\pi^2\sqrt{\epsilon}}\int_0^\infty d\tilde\kappa\,\tilde\kappa^4
\left(\frac{\sin\tilde\kappa\delta}{(\tilde\kappa\delta)^3}-\frac{\cos
\tilde\kappa\delta}{(\tilde\kappa\delta)^2}\right).
\eea
In these expressions we haven't written the remaining function of $\tilde
\kappa$ within the integrals.
The relation between the two cutoff factors is just that given in 
Ref.~\cite{Milton:2014psa}:
\be
\frac{d}{dx}\left(x\left[\frac{\sin x}{x^3}-\frac{\cos x}{x^2}\right]\right)
=-\frac{d^2}{dx^2}\frac{\sin x}x.\label{identity}
\ee

\section{Quadratic potential}
\label{appb}
Suppose the potential begins quadratically, that is, it is continuous,
with a continuous first derivative, but a discontinuous second derivative
at the edge, 
\be
\varepsilon(z)=1+\beta z^2.
\ee
We then easily find the  fundamental solution to first order in $\beta$:
\begin{subequations}
\bea
\left.\begin{array}{c}F^E(z)\\G^E(z)
\end{array}\right\}&=&e^{\mp\kappa z}\left[1\mp
\frac{\beta\zeta^2 z}{4\kappa^3}\left(1\pm \kappa z+\frac23(\kappa z)^2\right)
\right],\\
\left.\begin{array}{c}F^H(z)
\\G^H(z)\end{array}\right\}&=&e^{\mp\kappa z}\left[1\mp
\frac{\beta z}{4\kappa^3}\left((\zeta^2-2\kappa^2)\pm(\zeta^2-2\kappa^2)
 \kappa z+\frac23\zeta^2(\kappa z)^2\right)
\right].
\eea
\end{subequations}
Here we again note that the terms proportional to $\zeta^2$ are identical.
The Wronskians of the solutions are
\begin{subequations}
\bea
\alpha^E=w^E&=&2\kappa+\frac{\beta\zeta^2}{2\kappa^3},\\
\alpha^H=\frac{w^H(z)}{\varepsilon^H(z)}&=&2\kappa+\frac\beta{2\kappa^3}
(\zeta^2-2\kappa^2).
\eea
\end{subequations}

First consider the bulk divergences.  The identity (\ref{bulkstred}) still
holds with the potential $\alpha\zeta^2\epsilon z$ here replaced by $\beta
\zeta^2 z^2$, so it is straightforward to compute in the plasma model,
where $\beta\zeta^2=\beta_0$ is a constant,
\be
T_{zz}^{b,E}=\frac1{2\pi^2\delta^4}-\frac{\beta_0z^2}{8\pi^2\delta^2}
-\frac{(\beta_0 z^2)^2}{32\pi^2}\ln\delta,
\ee
which is just as expected from the WKB analysis of Ref.~\cite{Milton:2016sev}.
The divergent terms are again the same
for the corresponding TM contributions. 
And for the energy density,  with temporal splitting
\be
T_{00}^{b,E}=\frac{3}{2\pi^2\delta^4}-\frac{\beta_0 z^2}{8\pi^2\delta^2}
+\frac{\beta_0}{24\pi^2}\ln\delta,
\ee
again as expected. 
  Although for the TM part a singularity emerges in the
$\zeta$ integration once again, the first two terms here are reproduced.

 For the nondispersive, constant $\beta$, case
we obtain results precisely analogous to those in Eqs.~(\ref{tbezz}),
(\ref{tbhzz}) and Eq.~(\ref{tbe00}):
\be
T^{b,E,H}_{zz}\sim \frac{1+\frac32\beta z^2}{2\pi\delta^4},\quad
T^{b,E,H}_{00}\sim \frac{3+\frac92\beta z^2}{2\pi\delta^4}.
\ee
For the scattering parts, we need the reflection coefficients:
\be
 \tilde R^E=\frac{\beta\zeta^2}{8\kappa^4},\quad
\tilde R^H=\frac{\beta}{8\kappa^4}(\zeta^2-2\kappa^2).
\ee
Then, for the normal-normal stress tensor and the energy density
we obtain terms which are less singular toward
the edge than was the case for the linear potential for the non-dispersive
case:
\begin{subequations}
\bea
 T^{s,E}_{zz}&\sim&-\frac{\beta^2}{640\pi^2}\ln z,\quad
 T^{s,H}_{zz}\sim-\frac{43\beta^2}{1920\pi^2}\ln z,\\
T^{s,E}_{00}&\sim&\frac\beta{960\pi^2z^2},\quad
T^{s,H}_{00}\sim-\frac{3\beta}{320\pi^2z^2}.
\eea
\end{subequations}
Notice that the ratios of the $zz$ components are 43/3, while the energy
densities are in the ratio $-9$, exactly as in the linear case, which reflects
the fact that the angular integrations over $\cos^2\theta=\zeta^2/\kappa^2$ are
the same.

\end{document}